\documentclass[aps,pra,twocolumn,superscriptaddress,showpacs]{revtex4}
\usepackage{graphicx}
\usepackage{amsmath}
\usepackage{epsfig}
\usepackage{epstopdf}
\usepackage{subfigure}
\pdfoutput=1 

\begin{document}

\title{Quantum Trajectory Analysis of the Two-Mode Three-Level Atom Microlaser}

\author{Tarek A. Elsayed}
%\email{physics.reality@gmail.com}
\email{T.elsayed@thphys.uni-heidelberg.de, Tarek.Ahmed.Elsayed@gmail.com}
\affiliation{Electronics Research Institute, Dokki, Giza 12622, Egypt}
\affiliation{Institute of Theoretical Physics, University of Heidelberg, Philosophenweg 19, 69120 Heidelberg, Germany}

\author{Abdulaziz Aljalal}
\email{Aljalal@kfupm.edu.sa}
%\thanks{}
\affiliation{King Fahd University of Petroleum and Minerals, 31261 Dhahran, Saudi Arabia}


%\date{\today}

\begin{abstract}
We consider a single atom laser (microlaser) operating on three-level atoms interacting with a two-mode cavity. The quantum statistical properties of the cavity field at steady state are investigated by the quantum trajectory method which is a Monte Carlo simulation applied to open  quantum systems. It is found that a steady state solution exists even when the detailed balance condition is not guaranteed. The differences between a single mode microlaser and a two-mode microlaser are highlighted. The second-order correlation function \(g^{(2)} (\tau )\) of a single mode is studied and special attention is paid to the one-photon trapping state, for which a simple formula is derived for its correlation function. We show the effects of the velocity spread of the atoms used to pump the microlaser cavity on the second-order correlation function, trapping states, and phase transitions of the cavity field.

\end{abstract}
\pacs{42.50.Pq, 42.50.Ar, 42.50.Lc}
\keywords{Microlaser, Quantum Trajectory Method, Coherence Function, Field statistics}

\maketitle
\section{Introduction}
The single atom laser (microlaser) has been used successfully in the last decade for studying the quantum nature of the atom-field interaction in the optical wavelength regime \cite{An,Mad,Choi,pra2006,pra2009}. Following the same principles of the single atom maser (micromaser), the microlaser operates by pumping a high finesse optical cavity by a low-density beam of excited resonant atoms. Due to the coherent nature of the interaction between the atoms and the cavity field, some properties that are statistically averaged out in the conventional laser are evident in the microlaser \cite{book}. As a platform to investigate other unusual features of the laser operation such as multiple threshold operation \cite{pra2006}, the microlaser has become one of the most fundamental systems in quantum optics \cite{pra2009}. 

 Novel features of the non-classical state of radiation in the microlaser cavity and the emitted beam of photons such as sub-poissonian photon statistics and anti-bunching have been recorded \cite{Choi,pra2006}.
An important tool to investigate these non-classical properties of the emitted and stored radiation is the second-order correlation function \(g^{(2)} (\tau )\). 

Our main focus in the  current work is to study the statistical properties, including the second-order correlation function,  of the cavity field of an interesting type of microlaser that utilizes bi-modal cavities pumped by multi-level atoms. \\

Bimodal atom-field interaction has received considerable interest in  atomic physics and quantum optics \cite{Mess}. A two-mode microlaser has indeed shown theoretical promise for the novel feature of lasing without population inversion \cite{Fam95}. In this last reference, Kien et. al.  have analyzed a bimodal laser pumped by three-level atoms in \(\Lambda\)-configuration analytically. In this study, we use the quantum trajectory method \cite{Dal,Dum,Cres,Car} to analyze a similar two-mode microlaser operated by pumping a doubly resonant optical cavity by excited atoms in \(\Lambda\)-configuration  characterized by two lasing transitions and strongly coupled to the cavity. The atoms interact resonantly with the two independent modes of the cavity that compete for the gain contribution \cite{Fam95}. In practice, the atoms are produced by an oven and the velocities of the atoms obey a thermal velocity distribution. The atoms then are allowed to pass through a velocity selector to unify the speeds of the atoms and consequently their interaction times with the cavity. Since the efficiency of the velocity selector is not perfect, atoms passing through the cavity still have a slight velocity variation. Contrary to the work in \cite{Fam95}, we consider that the operation is in the strong coupling regime and that atoms are injected in their upper state with different velocities. 
Including the variations of the atoms' velocities in the full analytical treatment of the microlaser is a difficult task \cite{Choi}. 
In this paper, we utilize the full power of the quantum trajectory method to include the variation of the atoms' velocities and investigate its effects on the statistical properties of the cavity field and specifically its second-order correlation function \(g^{(2)} (\tau )\).

We found that the correlation function of the mono-velocity two-level and three-level atom microlaser that is operated in a special state called \emph{the one photon trapping state} follows a simple exponential formula. Unlike the single mode two-level atom microlaser, the two-mode microlaser exhibits super-poissonian photon distribution in most of the operating conditions. We explain the origin of this difference in the text. We found also that  the  second-order correlation is  in general  robust against the  velocity  spread  of  the  atoms when the microlaser is operated far from the trapping states,  and that some residual correlation persists even at a wide velocity spread of the atoms. The impact of the atomic velocity spread on \(g^{(2)} (\tau )\) when the microlaser is operated at different trapping states is also investigated.
The results of our study should be illuminating for experiments conducted on this simple type of multimode laser. \\
 
 The rest of this paper is organised as follows. In Sec. II we review the theory of the two-mode microlaser and the correlation function of the field inside the microlaser cavity and elaborate on the interesting one-photon trapping state. In Sec. III we introduce the quantum trajectory method, apply it to the two-mode microlaser, and show the results of calculating the correlation function \(g^{(2)} (\tau )\) numerically and analytically. In Sec. IV, we show the results of including the variation of the atoms' velocities in the numerical simulation and elucidate its effect on the statistical properties of the microlaser field. We conclude our work in Sec. V and present suggestions for further investigations.

\section{Theory}
\begin{figure} 
\includegraphics[width=0.45\textwidth]{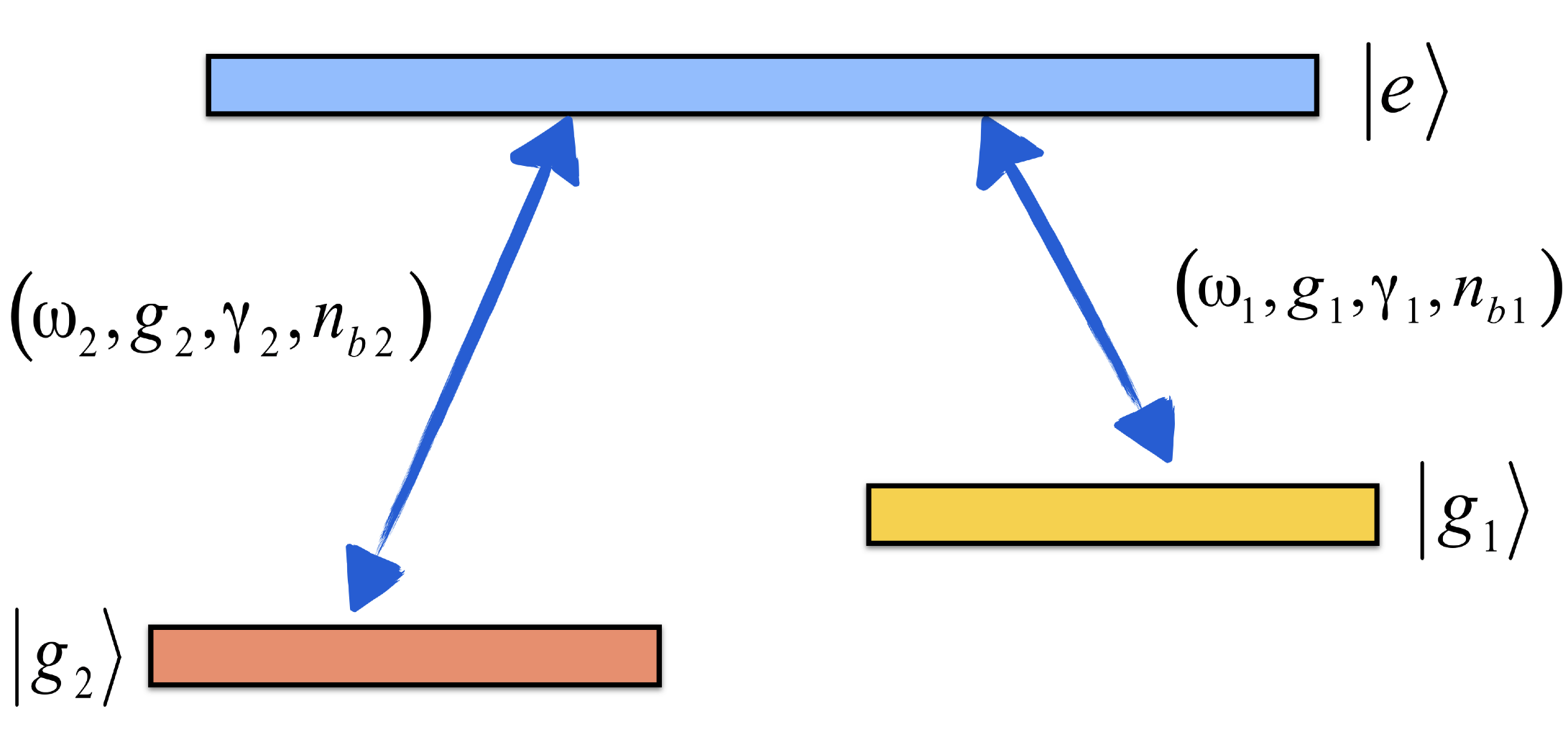}
\caption{ \label{levels}  A schematic diagram of the energy levels of a three level atom in $\Lambda$-configuration.}
\end{figure}

A schematic diagram showing the energy levels of the three-level atoms used to pump the bi-modal microlaser cavity is shown in Fig.~\ref{levels}. In the most general case, each mode has its own angular frequency \(\omega _\alpha\), coupling strength with the atom \(g_\alpha\), cavity decay rate \(\gamma _\alpha\), and mean number of thermal photons \(n_{b_\alpha}\)where \(\alpha = 1,2\). Unlike the micromaser, the mean number of thermal photons for a microlaser is typically zero; however, we include \(n_{b1}\), \(n_{b2}\) in the analytical part of this paper for the sake of generality. All the atoms are excited to the higher level \(\left| a \right\rangle\) before they enter the cavity. Atoms are statistically independent (have random arrival times) and their dwelling time inside the cavity is much shorter than the mean inter-atomic arrival times as well as the cavity damping time so that field dissipation is neglected during the atom-cavity interaction interval. The rate of atoms injection is assumed to be small enough such that, at most, only one atom exists inside the cavity \cite{Filipowicz}. Therefore, the mean number of photons inside the cavity is very small and nonlinear effects (e.g. coupling between the two modes as in \cite{pra2009}) are negligible. 

We assume that the detuning between the two mode frequencies is large, compared to the atom-field coupling strengths \(g_1 \) and \(g_2\), so that each mode interacts only with the respective atomic transition. Since the cavity is subjected to incoherent field decay, we are going to use density matrix formalism where the field is represented by its density matrix \(\rho\). The diagonal elements of the field density matrix \(P_{n_1 ,n_2 }\), which represent the joint probability to find \(n_1\) photons in mode 1 and \(n_2\) photons in mode 2, constitute a two dimensional matrix. In what follows, we summarize the procedure given in \cite{Fam94} to calculate the steady state solution of \(P_{n_1 ,n_2 }\). We are not interested in the off-diagonal elements of the density matrix in the current work. Changes to the density matrix are induced by two different and independent procedures, the atom field interaction and the field dissipation from the cavity. The change to the density matrix due to the interaction between the field and a single atom during the interaction time \(\tau_{{\mathop{\rm int}} }\) is represented by the super-operator \(F\left( \tau  \right)\) defined as \(
\rho (t_i  + \tau _{{\mathop{\rm int}} } ) = F\left( {\tau _{{\mathop{\rm int}} } } \right)\rho (t_i )
\). The change of the density matrix  due to the field dissipation is controlled by the Liouvillian super-operator \(\hat L_0\) given by \cite{scully}
\begin{widetext}
	\[
	\dot \rho|_{dissipation} = \hat L_0 \rho  =  - \frac{1}{2}\gamma _1 \left( {n_{b1}  + 1} \right)\left[ {a_1 ^ \dag  a_1 \rho  + \rho a_1 ^ \dag  a_1  - 2a_1 \rho a_1 ^ \dag  } \right] -\\ \frac{1}{2}\gamma _1 n_{b1} \left[ {a_1 a_1 ^ \dag  \rho  + \rho a_1 a_1 ^ \dag   - 2a_1 ^ \dag  \rho a_1 } \right]
\]
\begin{equation}
%\begin{split}
- \frac{1}{2}\gamma _2 \left( {n_{b2}  + 1} \right)\left[ {a_2 ^ \dag  a_2 \rho  + \rho a_2 ^ \dag  a_2  - 2a_2 \rho a_2 ^ \dag  } \right] -\\ \frac{1}{2}\gamma _2 n_{b2} \left[ {a_2 a_2 ^ \dag  \rho  + \rho a_2 a_2 ^ \dag   - 2a_2 ^ \dag  \rho a_2 } \right]
,	
%\end{split}
\end{equation}
\end{widetext}

where \((a_1 ^ \dag  ,a_1 ) \) and \((a_2 ^ \dag  ,a_2 ) \) are the field operators of the two modes, and \(n_{b1} \), \(n_{b2}\) are the mean number of thermal photons in the first mode and second mode,  respectively. 
It can be shown \cite{meystre} that under the aforementioned conditions and for the random arrival of atoms, which is a Poisson process, the master equation controlling the change of the density matrix \(\rho\) is given by
\begin{equation}
\dot \rho  = R\left[ {F(\tau _{{\mathop{\rm int}} } ) - 1} \right]\rho  + L_0 \rho, 
\end{equation}
where R is the rate of the atoms injection.		
At steady state, we have
\begin{equation}\label{master1}
\dot \rho  = 0,
\end{equation}
\begin{equation}\label{master2}
R\left[ {1 - F(\tau _{{\mathop{\rm int}} } )} \right]\rho  = L_0 \rho,
\end{equation}
which simply means that, the change per unit time in the field density matrix due to the atom field interaction is exactly compensated by the change due to field dissipation from the cavity. In other words, the net change per unit time of the field density matrix is zero. In Appendix A, we show how $F(\tau _{{\mathop{\rm int}} } )$ can be found by  solving the Schrodinger's equation during the transit time \cite{Fam94}. The result of this procedure in terms of the diagonal element \(P_{n,m}\)  is given by
\begin{widetext}
	\[	P_{n,m} (\tau _{{\mathop{\rm int}} } ) = P_{n,m} (0)cos^2 \left[ {\lambda \left( {n,m} \right)\tau _{{\mathop{\rm int}} } } \right] + g_1^2 n P_{n - 1,m} (0)\frac{{\sin ^2 \left[ {\lambda \left( {n - 1,m} \right)\tau _{{\mathop{\rm int}} } } \right]}}{{\lambda ^2 \left( {n - 1,m} \right)}}
\]
\begin{equation}\label{ftint}	 
+ g_2^2 m P_{n,m - 1} (0)\frac{{\sin ^2 \left[ {\lambda \left( {n,m - 1} \right)\tau _{{\mathop{\rm int}} } } \right]}}{{\lambda ^2 \left( {n,m - 1} \right)}},
\end{equation}
where $\lambda \left( {n,m - 1} \right)$ is defined in (\ref{lambda}).
\end{widetext}

 The effect of spontaneous emission is usually  neglected in the  analysis of  cavity QED experiments operated in the strong coupling regime \cite{An,thesis,Fangyen,pra2006}. The reason for this neglect is that during the transit time, the coherent interaction between the atom and the cavity field is the dominant interaction. It is shown in  Appendix A that when the coupling constant between the atom and cavity field is much higher than the spontaneous decay rate of the excited state of the atom $\Gamma_a $, i.e., \(g >> \Gamma_a \), in addition to short transit times compared to the lifetime of the excited level, $\Gamma_a \tau_{int} <<1$, the effect of spontaneous emission is indeed  negligible.

By substituting $F(\tau _{{\mathop{\rm int}} } )$ into (\ref{master1}) and (\ref{master2}) and rewriting it in terms of the diagonal elements of the density matrix, we get

\begin{widetext}
\[\dot P(n_1 ,n_2 ) = 0 = R\{- \sin ^2 \left[ {\lambda (n_1 ,n_2 )\tau _{{\mathop{\rm int}} } }\right]P(n_1 ,n_2 ) + g_1^2 n_1 \frac{{\sin ^2 \left[ {\lambda (n_1  - 1,n_2 )\tau _{{\mathop{\rm int}} } } \right]}}{{\lambda ^2 (n_1  - 1,n_2 )}}P(n_1  - 1,n_2 ) \]
\[+ g_2^2 n_2 \frac{{\sin ^2 \left[{\lambda (n_1 ,n_2  - 1)\tau _{{\mathop{\rm int}} } } \right]}}{{\lambda ^2 (n_1 ,n_2  - 1)}}P(n_1 ,n_2  - 1)\}+\]
\[\gamma _2 (n_{b2}  + 1)\left[ {(n_2  + 1)P(n_1  ,n_2+1 ) - n_2 P(n_1 ,n_2 )} \right] + \gamma _2 n_{b2} \left[ {n_2 P(n_1 ,n_2-1 ) - (n_2  + 1)P(n_1 ,n_2 )}\right]+\]
\begin{equation}\label{ME}
 \gamma _1 (n_{b1}  + 1)\left[ {(n_1  + 1)P(n_1  + 1,n_2 ) - n_1 P(n_1 ,n_2 )} \right] + \gamma _1 n_{b1} \left[ {n_1 P(n_1  - 1,n_2 ) - (n_1  + 1)P(n_1 ,n_2 )}\right].
\end{equation}
\end{widetext}

\begin{figure}
\includegraphics[width=0.45\textwidth]{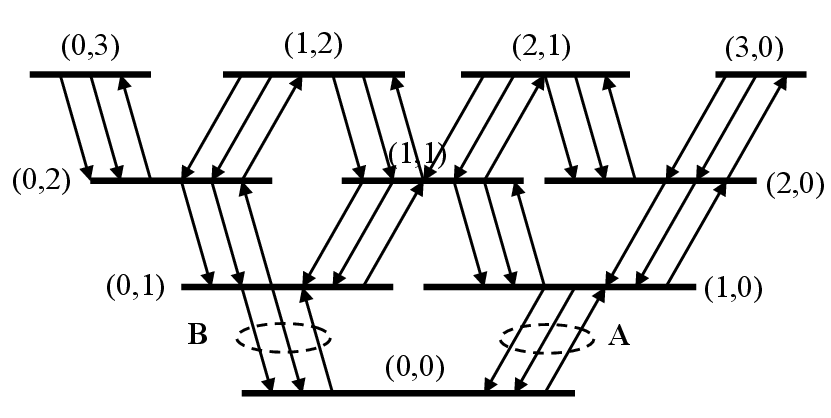}
\caption{ \label{levels2} A schematic diagram of the lowest energy states of the field inside a bi-modal cavity. The arrows represent the probability flow between adjacent energy states. Each state is denoted by the number of photons in the first and second mode, respectively.}
\end{figure}

 The previous equation contains the probability flow terms between any two successive cavity field energy levels as shown in Fig. ~\ref{levels2}.

Since the net probability flow from level (0,0) is zero at steady state, we conclude that the sum of the probability flow terms A and B is zero. Under the symmetric operation of the microlaser  defined by
\[
g_1  = g_2 ,{\rm{  }}\gamma _1  = \gamma _2 ,{\rm{  }}n_{b1}  = n_{b2}, 
\]
we can conclude that the probability flow bundles A and B are equal, and hence each of them will be identically zero. This defines the detailed balance condition which requires that transitions between any two states take place with equal frequency in either direction at equilibrium \cite{Thomsen}.
The equality of the coupling constants between the atomic transitions and the non-degenerate cavity modes can be realized experimentally by proper selection of the volumes of the two modes supported by two interlaced cavities, since the coupling constant is inversely proportional to the square root of the mode volume. This condition has been utilized in several studies \cite{Fam95,Fam94,comment1,comment2,comment3,comment4}.

 By induction, one can conclude that the three-term probability flow between any two levels in (\ref{ME}) is zero and we obtain \cite{Fam94}:
\begin{widetext}
\begin{equation}
	Rg^2 n_1 \frac{{\sin ^2 \left[ {\lambda (n_1  - 1,n_2 )\tau _{{\mathop{\rm int}} } } \right]}}{{\lambda ^2 (n_1  - 1,n_2 )}}P(n_1  - 1,n_2 ) = \gamma (n_b  + 1)\left[ {n_1 P(n_1 ,n_2 )} \right] - \gamma n_b \left[ {n_1 P(n_1  - 1,n_2 )}\right],
\end{equation}
\begin{equation}
	Rg^2 n_2 \frac{{\sin ^2 \left[ {\lambda (n_1 ,n_2  - 1)\tau _{{\mathop{\rm int}} } } \right]}}{{\lambda ^2 (n_1 ,n_2  - 1)}}P(n_1 ,n_2  - 1) = \gamma (n_b  + 1)\left[ {n_2 P(n_1 ,n_2 )} \right] - \gamma n_b \left[ {n_2 P(n_1 ,n_2  - 1)}\right],
\end{equation}
which can be solved together to get 

\begin{equation}\label{p-analytica}
	P(n_1 ,n_2 ) = P(0,0)\left( {\frac{R}{{\gamma (n_{b}  + 1)}}} \right)^{n_1+n_2 } \prod\limits_{k = 1}^{n_1  + n_2 } {\left( {\frac{{\gamma n_b }}{R} + \frac{1}{{k + 1}}\sin ^2 \left[ {g\tau _{{\mathop{\rm int}} } \sqrt {k + 1} } \right]}\right)}, 
\end{equation}
\end{widetext}

where \(P(0,0)\) is obtained from the normalization condition\(\sum\limits_{n_1 ,n_2 } {P(n_1 ,n_2 )}  = 1\). The probability of finding n photons in one mode regardless of the number of photons in the other mode, \(P_\alpha  (n)\)
is given by	
\begin{equation}
	\begin{array}{l}
	 P_1 (n) = \sum\limits_{n_2 } {P(n_1 ,n_2 )},  \\ 
	 P_2 (n) = \sum\limits_{n_1 } {P(n_1 ,n_2 )}.  \\ 
	 \end{array}
\end{equation}
We can see easily that \(P_1 (n) = P_2 (n)\), which is a signature of the symmetric operation of the microlaser. In Fig. \ref{pn} , we plot \(P_{\rm{\alpha }} \left( n \right)\) for \(n_b  = 0.1,\) \({\rm{ }}\frac{R}{\gamma } = 50\) and \( g\tau _{{\mathop{\rm int}} }\)  = 0.8, 0.9 and 1. Note that for a realistic operation of the microlaser, the mean number of thermal photons $n_b$ is almost zero, but this would not affect the current discussion. Unlike the photon statistics of the single mode microlaser, we notice in the probability distribution of a single mode of the two-mode microlaser, \(P_{\rm{\alpha }} \left( n \right)\), the existence of regions of exactly flat distribution. The explanation for these flat regions is plausible:  

\begin{figure}
\includegraphics[width=0.45\textwidth]{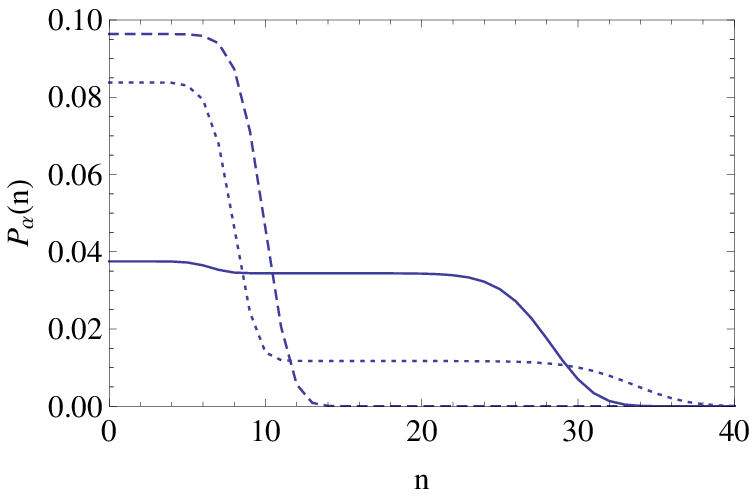}
\caption{ \label{pn} The photon number probability distribution \(P_{\rm{\alpha }} \left( n \right)
\) for mode 1, calculated for \(n_b  = 0.1\), \({\rm{ }}\frac{R}{\gamma } = 50,\) \({\rm{ }}g\tau _{{\mathop{\rm int}} }\)  = 0.8 (dashed), 0.9 (dotted), 1.0 (solid).}
\end{figure}

The semi-classical rate equation at steady state of the three-level atom microlaser involves the total number of photons in the two modes (n+m) and reads:

\begin{equation}\label{semi-eqn}
	R \ {\rm{ s}}in^2 \left[ {g\tau _{{\mathop{\rm int}} } \sqrt {n + 1 + m + 1} } \right] = \gamma \left[ {n + m} \right].
\end{equation}

\begin{figure}
\includegraphics[width=0.4\textwidth]{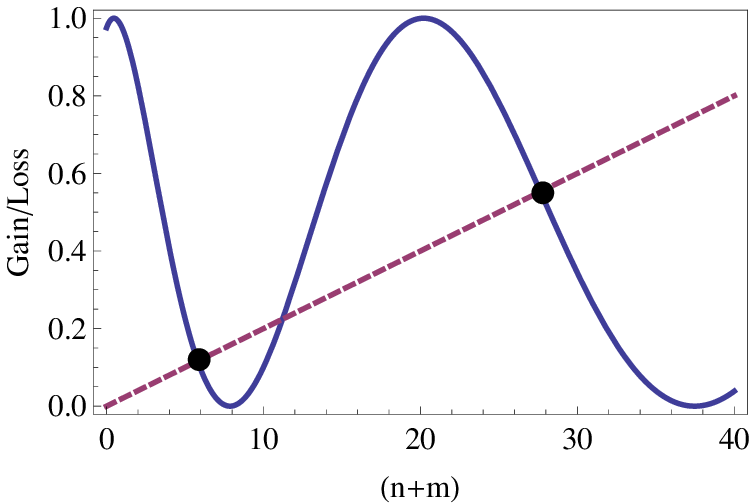}
\caption{ \label{semi} The dimensionless gain (solid) and loss (dashed) parts of the semi-classical rate equation expressed as $\ {\rm{ s}}in^2 \left[ {g\tau _{{\mathop{\rm int}} } \sqrt {n + 1 + m + 1} } \right]$ and $\gamma /R \left[ {n + m} \right]$ respectively for \(g\tau _{{\mathop{\rm int}} }  = 1 and {\rm{ }}R/\gamma = 50.\)
}
\end{figure}

 Stable solutions of (\ref{semi-eqn}), which correspond to the black intersection points in Fig. \ref{semi}, specify steady state values for (n+m). Since each value of the solutions for (n+m) can be formed by different combinations of n and m and all these combinations are equally probable, it follows that the probability distribution \(P_{\rm{\alpha }} \left( n \right)\) has a flat probability regions corresponding to all the different combinations of n and m. These flat regions are higher for lower n and steps down for larger values of n in a staircase pattern since for every possible solution for (n+m) the possible values of n and m start from zero, while the high values of n and m are accessible only for large values of the solutions of (n+m).

It is remarkable that these flat probability sections are a distinctive property of two-mode or higher microlasers.  They are responsible for a wider probability distribution than the poissonian distribution, i.e., super-poissonian distribution.  In the more general case, when the coupling constants between the atoms and the two modes are not identical, \(g_1  \ne g_2\) , the semi-classical rate equation will be as follows,
\begin{equation}
	R\ {\rm{ }}\sin^2 \left[ {\tau _{{\mathop{\rm int}} } \sqrt {g_1^2 \left( {m + 1} \right) + g_2^2 \left( {n + 1} \right)} } \right] = \gamma \left[ {n + m} \right].
\end{equation}
 It is evident that in this case, which was not considered in \cite{Fam94}, no flat regions will appear in \(P_{\rm{\alpha }} \left( n \right)\) since we have a single n and m that satisfy the equation. This prediction will be confirmed shortly after applying the quantum trajectory method to the microlaser problem.

\subsection{Second-order correlation}

The steady state of the microlaser, as implied above, does not mean that the field inside the cavity of a certain microlaser setup has a constant radiation intensity since each atom will induce a change in the field, and the field decay between two atoms induces further changes. In this part, we examine the correlation between these fluctuations in the cavity field, or more precisely, fluctuations in the number of photons inside the cavity represented by the second-order correlation function
\(g^{(2)}(\tau )\). Measurement of the correlation function of the cavity field may provide direct evidence of the quantized nature of light by detecting distinct correlation effects of the quantum field such as photon anti-bunching \cite{loudon}.	The second-order correlation function of the quantized field inside the cavity, \(g^{(2)} (\tau )\), is proportional to the probability of finding a pair of photons inside the cavity at steady state separated by a time \(\tau\) \cite{Car}, regardless of what happens to the photon number during this time interval, and is defined as: 
\begin{equation}
	g^2 (\tau ) = \frac{{\left\langle {a^ \dag  (0)a^ \dag  (\tau )a(\tau )a(0)} \right\rangle }_{ss}}{{\left\langle n \right\rangle_{ss} ^2 }}.
\end{equation}

The correlation function of a micromaser cavity field can be calculated analytically as shown by Quang \cite{Quang} by starting with a density matrix \(\tilde \rho (0)\), conditioned on the act of detecting and annihilating a photon from the cavity field:
\[
\tilde \rho (0) = \frac{{a\rho _{ss} a^ \dag  }}{{Tr\left[ {a\rho _{ss} a^ \dag  } \right]}},
\]
where $\rho _{ss}$ is the steady state solution of the master equation.
 We evolve this conditional density matrix by the micromaser master equation. The correlation function at time \(\tau\) will be proportional to the mean number of photons inside the cavity at time 
\(\tau\) as calculated by the evolved conditional density matrix \(\tilde \rho (\tau )\): 
\[
g^{(2)} (\tau ) = \frac{{Tr\left[ {a^ \dag  a\tilde \rho (\tau )} \right]}}{{\left\langle n \right\rangle}}.
\]
The quantum regression theorem has been used to derive this method. To calculate the evolution of the conditional density matrix \(\tilde \rho (\tau )\), we solve the master equation by the fourth order Runge Kutta Method.  The other method of calculating 
\(g^{(2)} (\tau )\) mimics what is done experimentally and is based on calculating the correlation between the times at which photons leak from the cavity \cite{aljalal}. We will use the second method in the next section to calculate \(g^{(2)} (\tau )\) numerically by the quantum trajectory method. 

\begin{figure}
{

\includegraphics[width=0.45\textwidth]{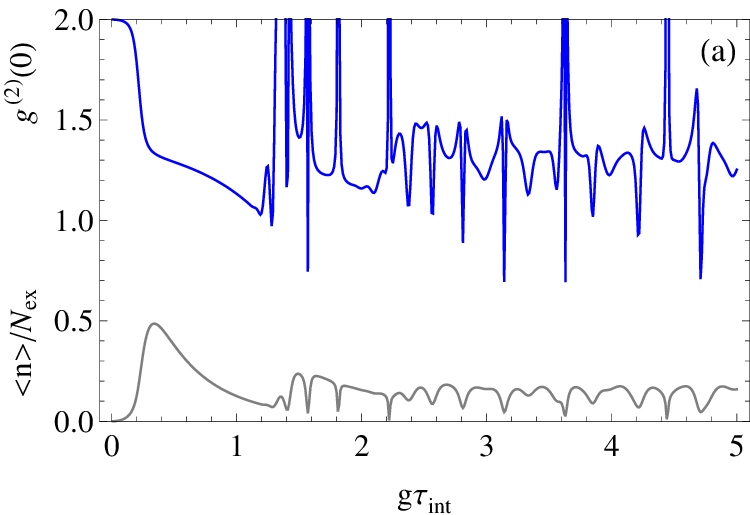}
 \label{g03} 
\hfill

\includegraphics[width=0.45\textwidth]{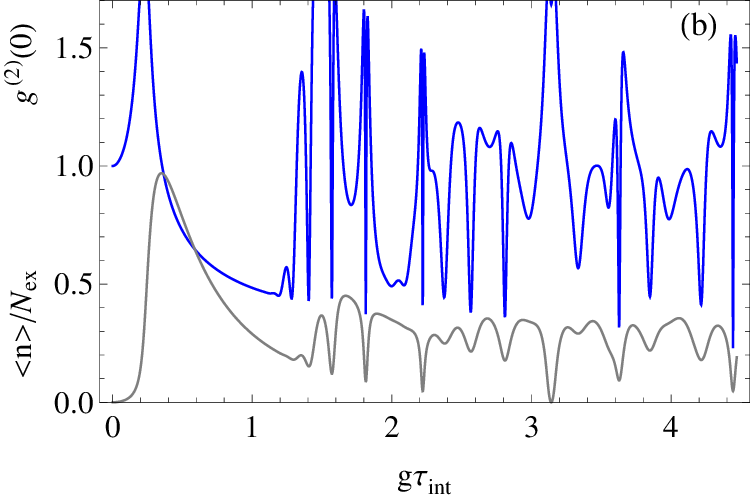}
 \label{g02} 
 
\caption{[Color Online] \label{g00} \(g^{(2)} (0)\)(blue)
and the normalized mean photon number (gray) plotted versus $g\tau _{{\mathop{\rm int}} }$ for a two-mode microlaser (a) and a single-mode microlaser (b).
}
}
\end{figure}

The equal-time correlation function \(g^{(2)} (0)\), which can easily be shown to satisfy  \cite{Quang}
\begin{equation}\label{g20eqn}
g^{(2)} (0)=\frac{\left \langle n^{2} \right \rangle-\left \langle n \right \rangle}{\left \langle n \right \rangle^{2}},
\end{equation}

 can be used as a measure of the variance of the photon number \(\sigma\)  defined by 
\[\sigma =\frac{\left \langle n^{2} \right \rangle-\left \langle n \right \rangle^{2}}{\left \langle n \right \rangle}.
\] It follows that 
\[g^{(2)} (0)=\frac{\sigma -1}{\left \langle n \right \rangle}+1\]
and hence when \(g^{(2)} (0)\) is greater than 1, the photon probability distribution will be wider than a poissonian distribution (super-poissonian) and vice versa.
When we plot  \(g^{(2)} (0)\), calculated for any of the two modes, versus the pumping parameter \(g\tau _{{\mathop{\rm int}} }\), as shown in Fig. \ref{g00}(a), we notice that \(g^{(2)} (0)\) is always higher than one except at the photon trapping states \cite{walther} characterized by the severe dips in \(g^{(2)} (0)\) and the normalized mean number of photons \(\left\langle n \right\rangle/{ N}_{ex } \) where \(\langle n \rangle= \sum_{ }^{ } n.{P }_{ \alpha  }(n)\) and \(
N_{ex}  = R/\gamma 
\). This means that photon number probability distribution \(P_{\rm{\alpha }} \left( n \right)\) is super-poissonian for most of the range of \(g\tau _{{\mathop{\rm int}} }\), a consequence of the flatness of \(P_{\rm{\alpha}}\left(n\right)\) explained earlier. This situation is not the same in the single mode microlaser due to the absence of this flatness. Fig. \ref{g00}(b) depicts the same graph for the single mode microlaser where the normalized mean number of photons and \(g^{(2)} (0)\) are plotted versus $g\tau _{{\mathop{\rm int}} }$.
It is evident that the photon number distribution exihibits both sub-poissonian and super-poissonian statistics in the single mode microlaser.

\begin{figure}
\includegraphics[width=0.45\textwidth]{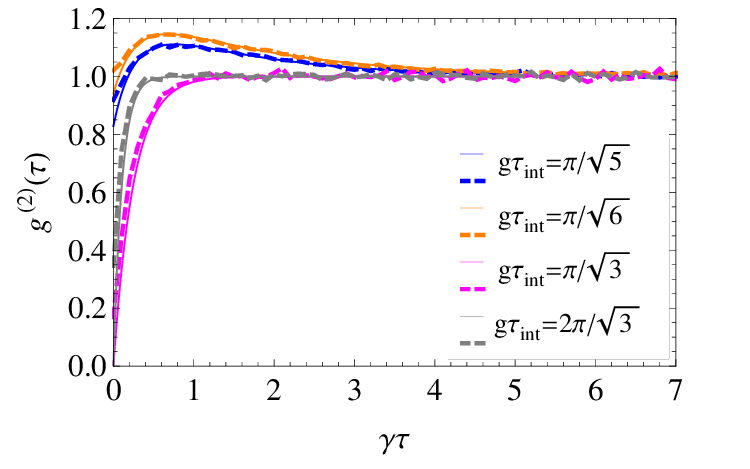}
\caption{ \label{G21}[Color Online] The second-order correlation function \(g^{(2)} (\tau)\) corresponding to \(g\tau _{{\mathop{\rm int}} }  = \frac{\pi }{{\sqrt 3 }}\) (magenta), $\frac{{2\pi }}{{\sqrt 3 }}$ (gray), \(\frac{\pi }{{\sqrt 5 }}\) (blue) and \(\frac{\pi }{{\sqrt 6 }}\) (yellow) calculated by QTM (dashed) and compared with the analytical method described above (solid).}
\end{figure}

 The rest of this section focusses on the correlation function at the trapping states defined by the condition 
\(g\tau _{{\mathop{\rm int}} } \sqrt {n + 1 + m + 1}  = k\pi\) for a certain (n+m). At these states, the probability of finding numbers of photons larger than (n+m) will be identically zero since the probability for each atom to emit a photon while interacting with a cavity field having m and n photons in the first and second mode, respectively, is proportional to \(\sin^2 \left[ {g\tau _{{\mathop{\rm int}} } \sqrt {n + 1 + m + 1} } \right]\). Fig. \ref{G21} shows \(g^{(2)} (\tau )\) calculated for the one, three, and four photon trapping states, compared with the numerical calculation by the quantum trajectory method. Severe anti-bunching characterized by \(g^{(2)} (\tau ) > g^{(2)} (0) \) is noticed for the one-photon trapping state. 

The  anti-bunching behavior of  the one photon  trapping state can be understood in  terms of  the necessary  time required between detecting a photon outside the cavity and re-pumping the cavity by an excited atom that deposits  another  photon  inside it.  This  dead  time  between  detecting  a  photon  and  re-pumping  the  cavity  is  responsible  for  the  photons  anti-bunching. This  situation  is  very similar to the anti-bunching of the fluorescence radiation emitted by a single atom where a 
dead time is unavoidable between the emission of a photon and re-exciting the atom.
According to the theory of resonance fluorescence of a single atom, the expression of the second-order correlation \(g^{(2)} (\tau )\) of the radiation scattered by a two-level atom driven by a continuous laser field is given by  \cite{Car,scully}
\[
g^{(2)} (\tau ) = 1 - \left( {\cos (\mu \tau)  + \frac{{3\gamma }}{{4\mu }}\sin (\mu \tau) } \right)e^{ - \frac{{3\Gamma }}{4}\tau}, 
\]
where $\Gamma$ is the spectral linewidth of the atom, or alternatively,  its spontaneous decay rate and $\mu$ is defined in terms of $\Gamma$ and the Rabi frequency \(\Omega _R \) by \(\mu  = \sqrt {\Omega _R ^2  - \frac{{\Gamma ^2 }}{{16}}}\).

\begin{figure}
\includegraphics[width=0.45\textwidth]{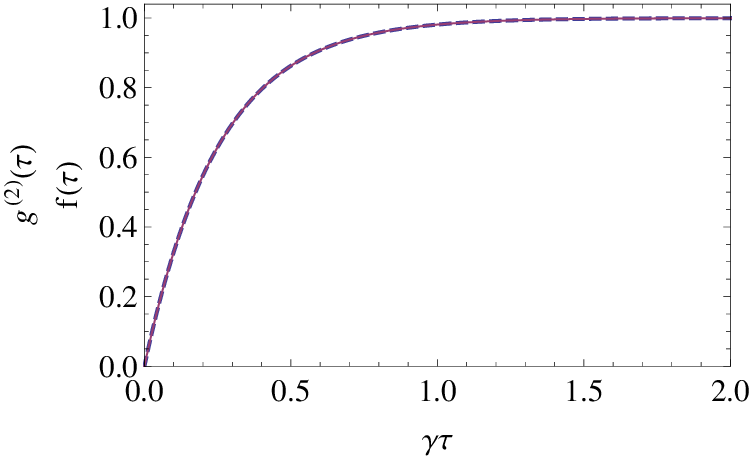}
\caption{ \label{G22} The second-order correlation function (dashed) \(g^{(2)} (\tau)\) for the one photon trapping state ($g\tau _{{\mathop{\rm int}} }  = \frac{\pi }{{\sqrt 3 }}$) fitted with the analytical function \(f(\tau )\) (solid).}
\end{figure}

The analogy between the two-level atom and the one-photon trapping state is clear. The cavity 
plays  the role of a  two-level system where  the  two  levels are either a photon stored  in 
the  cavity or not stored in the cavity. However, the  two systems differ in  the method of pumping. While  the  atom  undergoes Rabi oscillation by the continuous driving  laser field, the cavity is pumped by a 
stream of atoms arriving at random times and separated by relatively  large  intervals. Perhaps  this difference  is  the reason for not 
having the oscillatory behavior in the correlation function of the cavity field.
We tried to fit \(g^{(2)} (\tau )\) for this particular case with an analytical function and found that the function \(f(\tau )=1 - e^{ - \eta \tau }\) , where
\(\eta  = R\ \sin^2 \left[ {g\tau _{{\mathop{\rm int}} } \sqrt 2 } \right] + \gamma \), fits excellently with \(g^{(2)} (\tau )\) as shown in Fig. \ref{G22}. 
We found that the correlation function for the total number of photons (in the two modes) exhibits the same behavior and can be fitted with the same function. This behavior is not only pertinent to the two-mode microlaser, but also appears in the single-mode microlaser operating in its one photon trapping state characterized by \(\sin^2 \left[ {g\tau _{{\mathop{\rm int}} } \sqrt {2} } \right] = 0\). We found again that its correlation function fitted excellently with the analytical function 
\begin{equation}\label{ft}
	f(\tau ) = 1 - e^{ -\eta\tau},
\end{equation}
where
\(\eta  = R\ \sin^2 \left[ {g\tau _{{\mathop{\rm int}} } \sqrt 1 } \right] + \gamma. \)
A derivation of this relation is given in Appendix B.

\section{Applying the quantum trajectory method to solve the three-level atom microlaser}
A numerical method, basically a Monte Carlo simulation, applied to quantum systems to solve dissipative master equations was developed by three groups at approximately the same time in the early nineties \cite{Cres,Dum,Dal,Car}.  In this method, called quantum trajectory method (QTM), the observables of the system are obtained by averaging over many possible histories of the evolution of the system density matrix as a function of time. Each of these histories is called a trajectory, and its evolution is of a stochastic nature. Due to the statistical nature of quantum mechanics, taking the average over a large number of trajectories is equivalent to solving the master equation for this system. This concept is the essence of the quantum trajectory method. A certain trajectory can describe the stochastic evolution of the wavefunction or the density matrix of the open quantum system subjected to random quantum jumps representing its interaction with the reservoir. This method has a numerical advantage by reducing the computational power required to solve the master equation considerably, especially when the system consists of many degrees of freedom \cite{Dum}. Another advantage of the quantum trajectory method is the high level of control it allows over the parameters of the system. Including the variation of the atoms' speeds in the analytical solution of the master equation is a difficult task. However, in our numerical simulation we use the latter advantage to let each atom in any trajectory have a different velocity according to the velocity probability distribution in order to simulate what happens in reality. 
Instead of evolving the density matrix in each trajectory, we will evolve wavefunctions representing the state of the cavity field. The simplest wavefunction one can use to represent the quantized electromagnetic field is the number state \cite{Cres}. For this reason, and generalizing the quantum trajectory algorithm applied to the single mode micromaser developed by Pickles and Cresser in \cite{Cres}, we evolve two number states \(\left| m \right\rangle ,{\rm{ }}\left| {\rm{n}} \right\rangle \), simultaneously representing the deterministic number of photons in the two modes. In the most general case, where the two-mode micromaser cavity is maintained at a very low temperature, we can infer seven different events that may occur to the number state \(\left| {m,n} \right\rangle\) with the following probabilities:\newline
1- An atom emits a photon in the first mode, \(\left| {m,n} \right\rangle  \to \left| {m + 1,n} \right\rangle\), with probability \(
\frac{1}{A}R\left( {m + 1} \right)g_1^2 \ {\sin ^2 \left[ {\lambda (m,n)\tau _{{\mathop{\rm int}} } } \right]}/{{\lambda ^2 (m,n)}}.
\)\\
2-	An atom emits a photon in the second mode, \(\left| {m,n} \right\rangle  \to \left|{m,n+1}\right\rangle\), with probability \(
\frac{1}{A}R\left( {n + 1} \right)g_2^2 \ {\sin ^2 \left[ {\lambda (m,n)\tau _{{\mathop{\rm int}} } } \right]}/{{\lambda ^2 (m,n)}}\).\\
3-	A photon from the first mode leaks out from the cavity, \(\left| {m,n} \right\rangle  \to \left| {m - 1,n} \right\rangle\), with probability \(
\frac{1}{A}\gamma _1 (n_{b_1 }  + 1)m\).\\
4-	A photon from the second mode leaks out from the cavity, \(\left| {m,n} \right\rangle  \to \left| {m,n - 1} \right\rangle\), with probability \(
\frac{1}{A}\gamma _2 (n_{b2}  + 1)n\).\\
5-	A photon from the cavity walls is transferred to the cavity field of the first mode, \(\left| {m,n} \right\rangle  \to \left| {m + 1,n} \right\rangle\), with probability 
\(\frac{1}{A}\gamma _1 n_{b_1 } \left( {m + 1} \right)\).\\
6-	A photon from the cavity walls is transferred to the cavity field of the second mode, \(\left| {m,n} \right\rangle  \to \left| {m,n + 1} \right\rangle\), with probability 
\(\frac{1}{A}\gamma _2 n_{b_2 } \left( {n + 1} \right)\).\\
7-	An atom passes through the cavity without emitting any photons, \(\left| {m,n} \right\rangle  \to \left| {m,n} \right\rangle\), with probability \(\frac{1}{A}R\cos ^2 \left[ {\lambda (m,n)\tau _{{\mathop{\rm int}} } } \right],\) where
\begin{widetext}
 \[
A = R + \gamma _1 \left[ {(n_{b_1 }  + 1)m + n_{b_1 } (m + 1)} \right] + \gamma _2 \left[ {(n_{b_2 }  + 1)n + n_{b_2 } (n + 1)} \right].\]
\end{widetext}
The values of these probabilities can be derived by writing the master equation in the Lindblad form \cite{Cres}: 
\begin{equation}
	\dot \rho  =  - \frac{i}{\hbar }\left[ {\hat H,\rho } \right] + \sum\limits_m {\left[ {\hat C_m \rho \hat C_m ^ \dag   - \frac{1}{2}\left( {\hat C_m \hat C_m ^ \dag  \rho  + \rho \hat C_m ^ \dag  \hat C_m } \right)} \right]},
\end{equation}
where \(\hat C_m\) represents the jump operator representing event (m). The probability of event (m) at time $(t_1)$ is calculated by: 
\begin{equation}
p(m) = \frac{{\left\langle {\psi (t_1 )} \right|\hat C_m ^ \dag  \hat C_m \left| {\psi (t_1 )} \right\rangle }}{{\sum\limits_m {\left\langle {\psi (t_1 )} \right|\hat C_m ^ \dag  \hat C_m \left| {\psi (t_1 )} \right\rangle } }}
\end{equation}
and the  probability distribution function of the waiting times, $\tau$, between two successive jumps  is given by: \[p(\tau ) = e^{ - \int\limits_t^{t + \tau } {\sum\limits_m {\left\langle {\psi (t)} \right|\hat C_m ^ \dag  \hat C_m \left| {\psi (t)} \right\rangle dt} } }.\] For our system \(p(\tau )\) becomes: \[
p(\tau ) = e^{ - A\tau}.\]
The choice of number states as the propagated wavefunctions has the advantage that the effective Hamiltonian controlling the evolution of the wavefunction between jumps keeps the number states unchanged \cite{Cres}. For more on the quantum trajectory method, see \cite{Dum,Dal,Car}.

\begin{figure}
\includegraphics[width=0.45\textwidth]{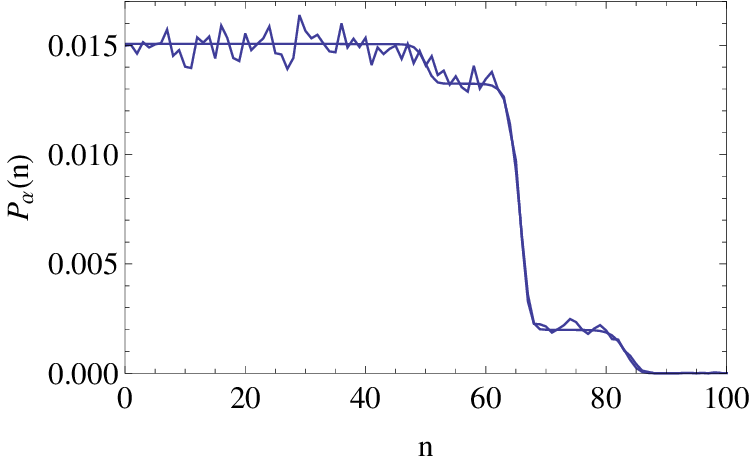}
\caption{ \label{pnqtm} The photon number probability distribution \(P_{\rm{\alpha }} \left( n \right)\) for mode 1 calculated numerically by QTM using 50000 trajectories and compared to the analytical solution for \(R/\gamma  = 200\) and \(\ {\rm{ }}g\tau _{{\mathop{\rm int}} } =0.3\).
}
\end{figure}

After generating many trajectories, we determine the diagonal elements of the density matrix by 
making a histogram over the final states \(\left| {m,n} \right\rangle\) of each trajectory. In Fig. \ref{pnqtm}, we show that the photon number probability distribution calculated by the quantum trajectory method for a microlaser operating at \(R/\gamma  = 200,\ {\rm{ }}g\tau _{{\mathop{\rm int}} }   = 0.3\) shows an acceptable accuracy as compared to the analytical solution (\ref{p-analytica}).

\begin{figure}
\includegraphics[width=0.45\textwidth]{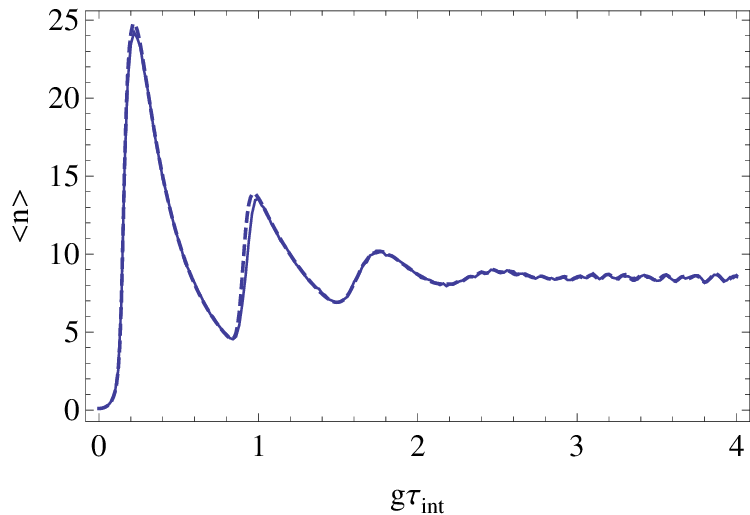}
\caption{ \label{meanqtm} The mean number of photons in mode 1 calculated numerically (solid) and compared to the analytical solution (dashed) for a micromaser operating at \(R/\gamma  = 50\) and nb=0.1. The number of trajectories in the numerical simulation is 10000 trajectories.}
\end{figure}

The accuracy becomes even better for lower values of \(R/\gamma\). Fig. \ref{meanqtm} shows the mean number of photons in one mode versus \(g\tau _{{\mathop{\rm int}} }\) obtained numerically and analytically. We used QTM to confirm that a microlaser operating in the non-symmetric mode, where the detailed balance condition does not apply (i.e., when transition frequencies between any two levels are not necessarily the same in both directions), does reach steady state by checking that the average density matrix calculated by QTM reaches a steady state. We show the evolution of $P(n,m)$ as it evolves from the vacuum state to the steady state for a microlaser operated in the non-symmetric mode at $g_1 \tau_{int}=9$ and $g_1 \tau_{int}=5$ in the movie included in the online supplementary material section.
 
\begin{figure}
\includegraphics[width=0.45\textwidth]{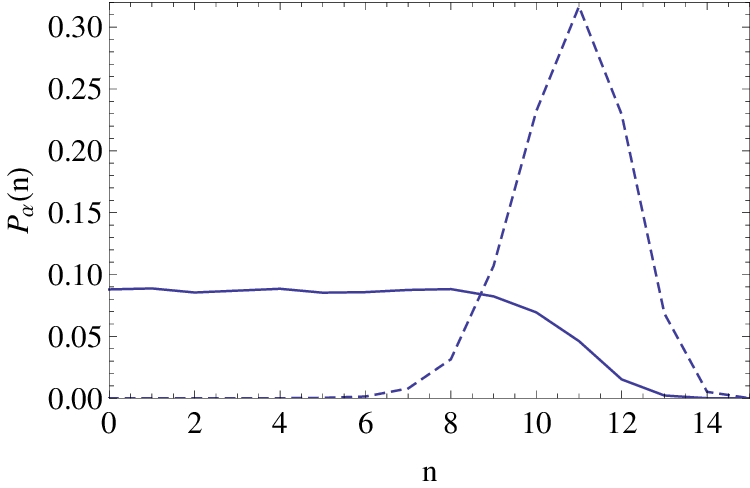}
\caption{ \label{nonsymmetric} The photon number probability distribution calculated numerically for \(g_{1}\tau _{int} =g_{2}\tau _{int}  = 0.8\) (solid) and \(g_{1}\tau _{int}  = 0.8,\) \(g_{2}\tau _{int}  = 0.5\) (dashed). The microlaser is operated at $ R/\gamma  = 100$. }
\end{figure}

\begin{figure}
\includegraphics[width=0.5\textwidth]{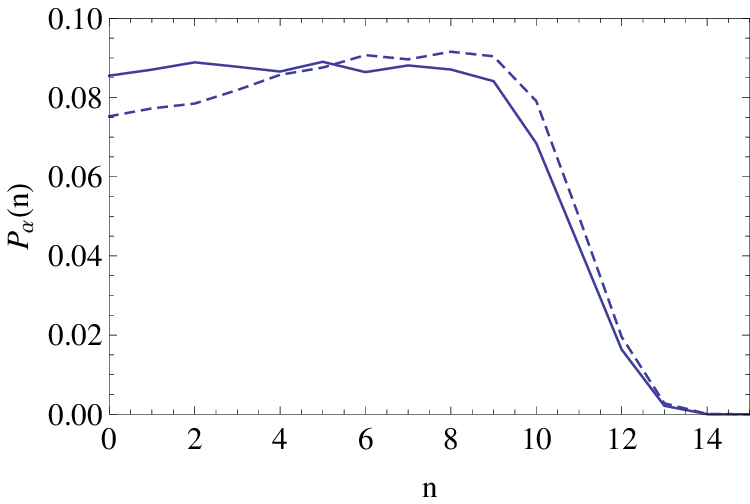}
\caption{ \label{nonsymmetric2} The photon number probability distribution calculated numerically for \(g_{1}\tau _{int} =g_{2}\tau _{int}  = 0.8\) (solid) and \(g_{1}\tau _{int}  = 0.8,\) \(g_{2}\tau _{int}  = 0.79\) (dashed). The microlaser is operated at $ R/\gamma  = 100$.}
\end{figure}

  In Fig. \ref{nonsymmetric} we see clearly that the flat regions disappear in the non-symmetric operation of the microlaser as predicted earlier. It turns out that this flatness in \(P_\alpha  \left( n \right)\) is very sensitive to the difference between the coupling constants of the two modes as shown in Fig. \ref{nonsymmetric2}. 

\begin{figure}
\includegraphics[width=0.45\textwidth]{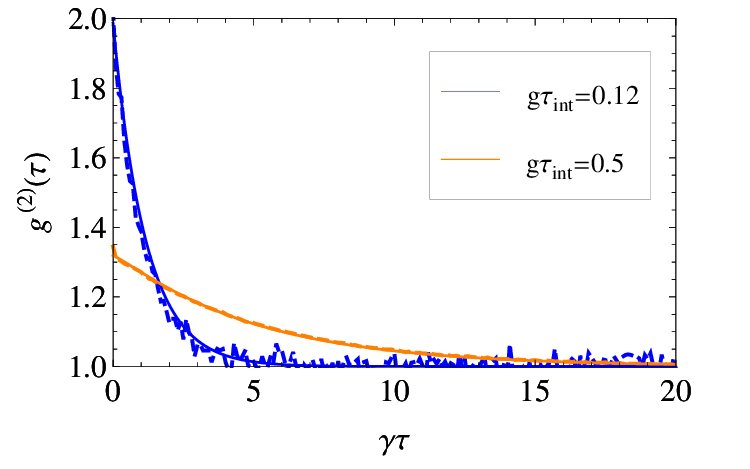}
\caption{ \label{bunching1} [Color Online] The second-order correlation function \(g^{(2)} (\tau )\) calculated for \(n_b = 0,\) \(\frac{R}{\gamma } = 10\),  \(g\tau _{{\mathop{\rm int}} }  = 0.12\) (blue) and  0.5 (orange) numerically (dashed) and analytically (solid).}
\end{figure}

\begin{figure}
\includegraphics[width=0.45\textwidth]{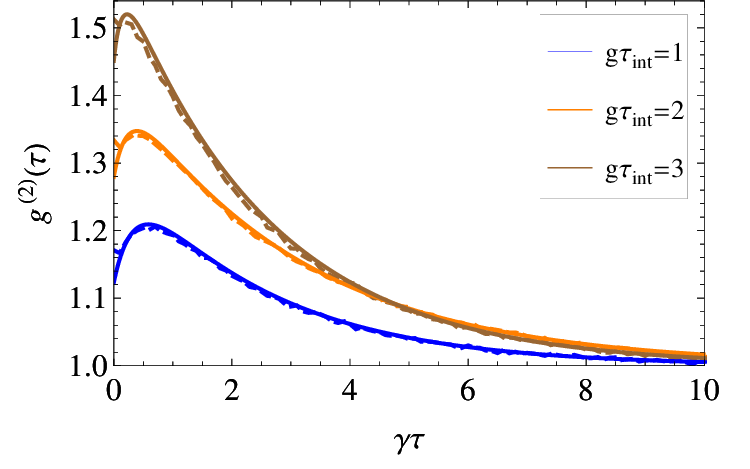}
\caption{ \label{bunching2}[Color Online] The second-order correlation function \(g^{(2)} (\tau )\) calculated for 
\(n_b  = 0, \frac{R}{\gamma } = 10,\) 	\(g\tau _{{\mathop{\rm int}} }  = 1 \) (blue), 2  (yellow) and  3 (brown)
numerically (dashed) and analytically (solid).}
\end{figure}

To calculate the second-order correlation function using the quantum trajectory method, we use a numerical method imitating the experimental method applied by Feld et al. \cite{aljalal,book}. 
In this experimental method, the coherence function of the microlaser is obtained by calculating the correlation between the times when photons are emitted out from the cavity and detected by the photo-detector. Numerically, we have full details about each trajectory including the times at which photons leak from the cavity. Therefore, \(g^{(2)} (\tau )\) is calculated by computing the correlation between these times in each trajectory and taking the average over all the trajectories. We have already shown in Fig. \ref{G21} the numerical calculation of \(g^{(2)} (\tau )\) compared with the analytical calculation for the trapping states where \(g^{(2)} (\tau )\) exhibits anti-bunching behavior. Fig. \ref{bunching1} shows the correlation function for two values of \(g\tau _{{\mathop{\rm int}}} \) where the cavity field exhibits bunching behavior. 
In Fig. \ref{bunching2}, we illustrate an interesting feature of \(g^{(2)} (\tau )\) where the correlation function exhibits a transient anti-bunching behavior before it decays monotonically to one. We should, however, point out the inability of this numerical method to predict accurately  the values of correlation function at very short times as evident in Fig.  \ref{G21} and \ref{bunching2}.

\section{EFFECT OF ATOMIC VELOCITY DISTRIBUTION ON THE STATISTICAL PROPERTIES OF THE MICROLASER}

\begin{figure}
\includegraphics[width=0.45\textwidth]{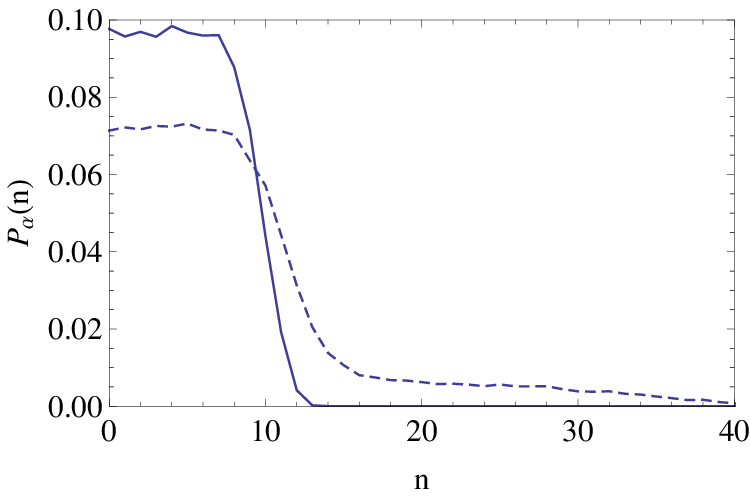}
\caption{ \label{pn20} The photon number probability distribution calculated numerically for the case of pumping the microlaser with a mono-velcoity atomic beam (solid) and an atomic beam having relative velocity spread of 20\% (dotted). The microlaser is operated at 
\(g\tau _{{\mathop{\rm int}} }\)=0.8 and  \(r/\gamma  = 50\).}
\end{figure}

As we mentioned briefly in the introduction, atomic velocity selectors are not perfect, and eventually the atoms passing through the microlaser cavity will have some velocity distribution. In this section, we illustrate the effect of this velocity spread of the atoms on the statistical properties of the field of one mode inside the microlaser cavity.
It might be expected that the velocity spread of the atoms would destroy the flat probability regions highlighted in the previous sections, but a quick look at Fig. \ref{pn} tells us that this expectation is not correct. In fact, the inclusion of a variety of interaction times will average the flat regions in the probability distribution corresponding to each value of \(\tau _{{\mathop{\rm int}}}\), and we end up with a persistent flat probability distribution whose width and height is a function of the relative atomic velocity spread as shown in Fig. \ref{pn20}. In this figure, we show that a relative velocity spread of 20\% maintains the flat regions in \(P_{\rm{\alpha }} \left( n \right)\) and consequently the super-poissonian character of the distribution.

\begin{figure}
\includegraphics[width=0.45\textwidth]{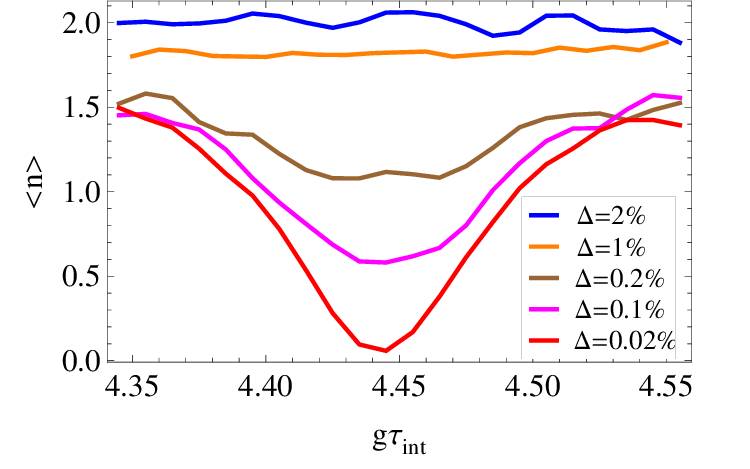}
\caption{ \label{trapping}[Color Online] The mean number of photons at a trapping state for different relative velocity spreads for a microlaser operating at \(R/\gamma  = 10\).}
\end{figure}

 A vacuum trapping state is a special trapping state where the cavity field is trapped at the vacuum state and occurs when the condition 
\(g\tau _{{\mathop{\rm int}} } \sqrt {1 + 1}  = n\pi\) is satisfied. This state, like other trapping states, is characterized by a sharp dip in the plot of the mean  number of photons. It is evident that the randomness in the interaction time \(\tau _{{\mathop{\rm int}} }\) will destroy this condition and remove the resonances from the microlaser behavior. We illustrate this behavior in Fig. \ref{trapping} where the mean number of photons in a vacuum trapping state is plotted for relative velocity distribution widths of  .02\%, .1\%, 0.2\%, 1\% and 2\%. It is notable that the vacuum trapping state is very sensitive to the velocity distribution width. 

\begin{figure}
\includegraphics[width=0.45\textwidth]{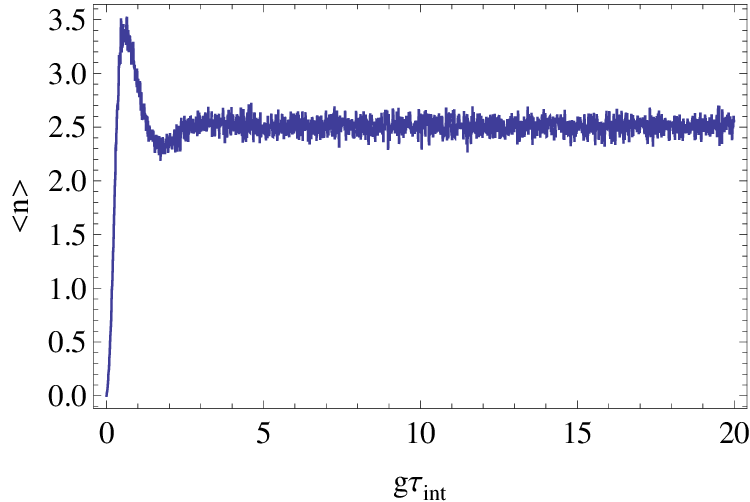}
\caption{ \label{mean60} The mean number of photons of the two-mode microlaser when relative velocity spreads of 60\% is included.}
\end{figure}

\begin{figure}
\includegraphics[width=0.45\textwidth]{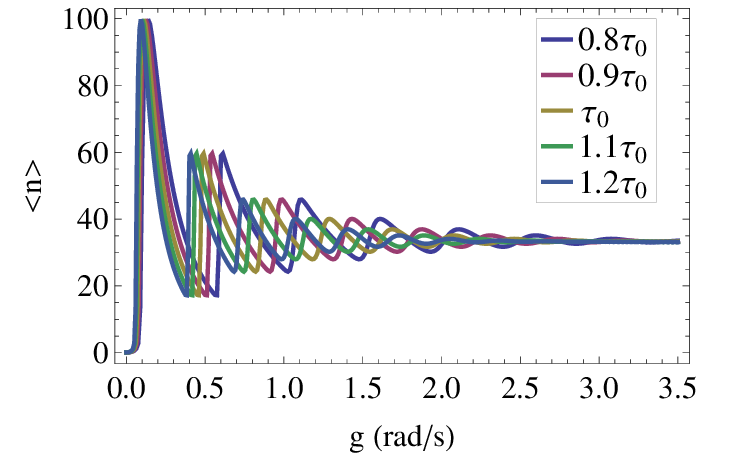}
\caption{ \label{differenttimes}
 [Color Online] The mean number of photons plotted versus the coupling constant \(g\) for different values of the interaction times to illustrate how a distribution of the interaction times (or equivalently atomic velocities) affect the microlaser phase transitions. \(\tau _0 \) is an arbitrary value of the interaction time and is equal to 1 s. The figure is for illustration and the values of g and \(\tau _0 \) are not realistic.}
\end{figure}

We noticed the existence of sharp peaks in the mean number of photons plotted versus 
\(g\tau _{{\mathop{\rm int}} }\) in Fig. \ref{meanqtm}. These peaks occur at the transitions between different stationary solutions of the semi-classical rate equation \cite{Filipowicz}, and we expect them to be induced by the fluctuations involved in the quantum system. We noticed, however, that including the spread in the atoms' velocities and hence increasing the randomness in the interaction times destroys these transitions starting from the transitions at large values of \(g\tau _{{\mathop{\rm int}} }\), which are very sensitive to the velocity broadening. This behavior is shown in Fig. \ref{mean60} where the mean number of photons is plotted versus \(g\tau _{{\mathop{\rm int}} }\) for a velocity distribution of width 60\%. We conclude from this observation that the system becomes more classical as more velocity fluctuations are introduced \cite{Filipowicz} and that noise in the pumping parameter destroys phase transitions in the micromaser \cite{Rekdal}. 
We can understand the immunity of the first transition to the velocity distribution and the fragility of the transitions at higher values of \(g\tau _{{\mathop{\rm int}} }\) by plotting the mean number of photons versus \(g\) for different values of the interaction times \(\tau _{{\mathop{\rm int}} }\). By taking the average of the different plots in Fig. \ref{differenttimes}, which correspond to a relative velocity spread of 40\%, we see the reason that the first phase transition is not much affected while the higher transitions are easily destroyed. 

\subsection{Effect of velocity spread on the correlation function}
\begin{figure}
\subfigure[  \( \) The microlaser is operated far from a trapping state, \(g\tau _{{\mathop{\rm int}} }  = 0.6\). Relative velocity spread \(\frac{{\Delta v}}{{v_o }} = 0\%,\) 20\% , 50\% and 100\%.]
{
\includegraphics [width=0.45\textwidth]{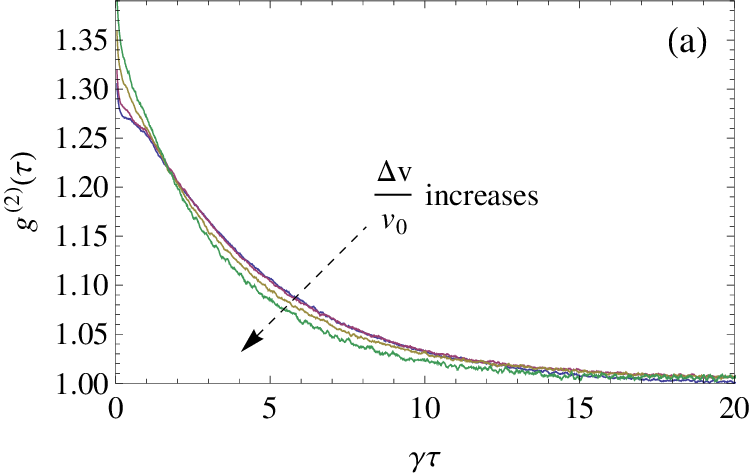}
\label{g2d1} 
}

\subfigure[  \( \) The microlaser is operated near a vacuum trapping state, \(g\tau _{{\mathop{\rm int}} }  = 1.994\pi /\sqrt 2\). Relative velocity spreads \(\frac{{\Delta v}}{{v_o }}\) = 0.02\%, 0.03\%, 0.04\%, 0.06\%, 0.08\%, 0.1\%, 0.2\%, 1\% from above to below respectively.]
{
\includegraphics[width=0.45\textwidth] {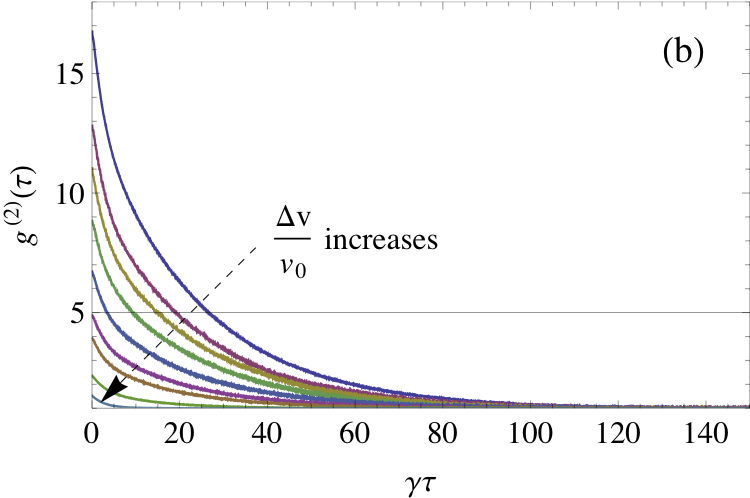}
\label{g2d2}
}
\caption{\label{g2d200} The second-order correlation function for different velocity spreads.}
\end{figure}

We have seen in the theory of the three-level two-mode microlaser that the photon statistics of the cavity field exhibits super-poissonian behavior for most of the range of the pumping  parameter except at some of the trapping states, where the field is anti-bunched.  We are going to show  the effect of the velocity spread on the second-order correlation function when there is a trapping state and when there is no trapping state.  Two regions are distinguished from Fig. \ref{g00}(a):  the first is the smooth region where \(g^{(2)} (0) < 2\); and the second one is the sharp peaks of \(g^{(2)} (0)\) at the vacuum trapping states where \(g\tau _{{\mathop{\rm int}} }\) is a multiple of \(\pi /\sqrt 2 \). 
For the first case, we observed that the correlation is immune to the atomic velocity spread and even a very broad velocity distribution does not reduce the field correlation substantially as shown in Fig. \ref{g2d1}. In this figure, the correlation due to mono-velocity atomic beam for a microlaser operating at 
\(g\tau _{{\mathop{\rm int}} }  = 0.6\) and \(R/\gamma  = 10\) is compared with the same microlaser pumped by an atomic beam having a spread of \(\frac{{\Delta v}}{{v_o }} = \ {\rm{ 20\% ,\ 50\% \ and \ 100\% }}\). While increasing the velocity spread changes the average velocity for the atoms and hence drifts the operating point of the microlaser slightly, what we want to emphasize in Fig. \ref{g2d1} is that the correlation is not affected much by the velocity spread.

 Finite correlation for practical atomic beams of relative velocity spreads up to 45\% has indeed been measured for the single-mode microlaser  \cite{Choi}.  For the second case in the vicinity of the  trapping states, \(g^{(2)} (0)\) is peaked because the number of photons inside the cavity is very small. The correlation function at these regions is strongly bunched, which means that whenever a few number of photons happen to exist inside the cavity, they will tend to leave the cavity together as a bunch of photons. In the extreme case of the vacuum trapping state, the correlation function diverges since the cavity has no photons. We noticed that when operating the cavity near a vacuum trapping state where \(g^{(2)} (0)\) is very large, the correlation is very sensitive to the velocity spread of the atoms and collapses very quickly until some residual correlation persists around a relative velocity spread of 1\%. In Fig. \ref{g2d2} we show the second-order correlation function of the cavity field \(g^{(2)} (\tau )\) for a microlaser operating close to this state (\(g\tau _{{\mathop{\rm int}} }  = 1.994\pi /\sqrt 2\)). In this figure, \(g^{(2)} (\tau )\) of a mono-velocity beam is shown in addition to velocity spreads \(\frac{{\Delta v}}{{v_o }}\) of 0.02\%, 0.03\%, 0.04\%, 0.06\%, 0.08\%, 0.1\%, 0.2\% and 1\%. The reason for the fragility of the correlation function near a vacuum trapping state is that the smallest distribution in the interaction times destroys the vacuum trapping state as we saw in Fig. \ref{trapping} and will introduce a small number of photons inside the cavity. These photons will cause the residual correlation mentioned above. Increasing the relative velocity spread beyond 1\%  does not affect this residual correlation considerably as in the first case (non-trapping states) where the correlation is very immune to the atomic velocity spread.

\begin{figure}
\subfigure[ \( \) The relative velocity spreads are \(\Delta {\rm{v/v_0 = 0}}{\rm{.01\%}}\), \(0.03\% \), \(0.05\%\),    \(0.07\% \), \(0.1\% ,\)  and 0.2\% from below to above respectively.]
{
\includegraphics [width=0.45\textwidth]{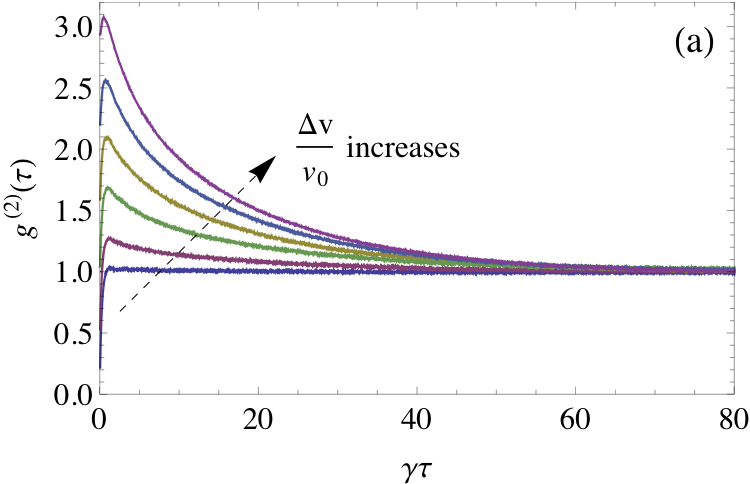}
\label{g2d3}
}
\subfigure[ \( \) The relative velocity spreads are \(\Delta {\rm{v/v_0 = 0}}.2\%\),  0.6\%,  1\%,  2\% and  20\% from above to below respectively.]
{
\includegraphics [width=0.45\textwidth]{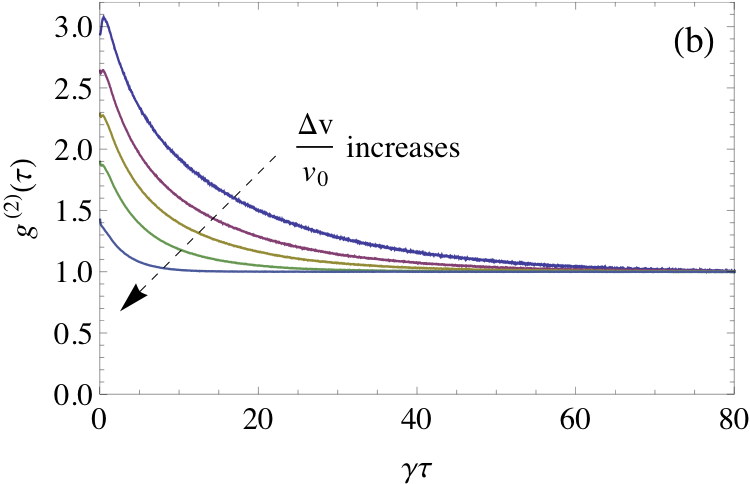}
\label{g2d4}
}
\caption{\label{g2d00} The second-order correlation function for a one-photon trapping state of a microlaser operated at \(g\tau _{{\mathop{\rm int}} }  = \pi/\sqrt{3}, R = 10.\)}

\end{figure}

On the other hand, when we investigated the effect of the atomic velocity spread on the correlation function for an operating point exhibiting anti-bunching, i.e., the one-photon trapping state, we found an interesting phenomenon. Adding more fluctuations in the atoms' velocities produces a peak in the correlation function near \(\tau  = 0\), and the field gradually shows more correlation  up to a relative velocity spread of \(\frac{{\Delta v}}{{v_o }} = 0.2\% \) as shown in Fig. \ref{g2d3}. The explanation for this weird behavior of noise-induced correlation is as follows: 
when the velocity spread is slightly increased starting from the mono-velocity case, there will be a very small probability of finding numbers of photons inside the cavity higher than one photon, i.e., a bunch of photons. Although the probability is very small compared to the probability of finding one photon or zero photon, its effect is overwhelming, and eventually the correlation function is dominated by these rare bunches of photons that can exist inside the cavity.

\begin{figure}
\includegraphics [width=0.45\textwidth]{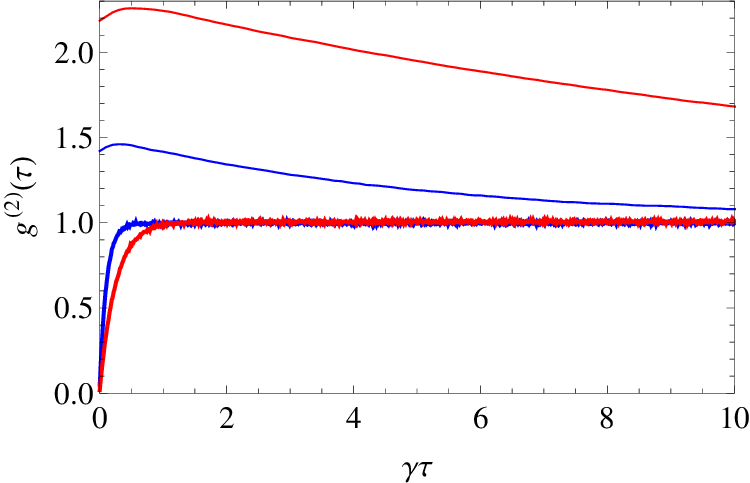}
\caption{\label{g2d5}[Color Online] The second-order correlation function for the total number of photons for the one-photon trapping states \(g\tau _{{\mathop{\rm int}} }  =\pi/\sqrt{3}\) (red) and \(2\pi/\sqrt{3}\) (blue) and relative velocity spreads \(\Delta {\rm{v/v_0 = 0\%}}\)  (thick) and  0.4\%  (thin).}
\end{figure}

We can explain the effect of the atomic velocity spread on the value of \(g^{(2)} (0)\)  quantitatively (see equation (\ref{g20eqn})) as follows:

For the one photon trapping state, we have
\(P(0),P(1) \ne 0\) and \(P(n) = 0\) for \(n > 1\) and hence \(\left\langle{n^2}\right\rangle=\left\langle n \right\rangle  = P(1)\),  making \(g^{(2)} (0) = 0\).  For the slightest velocity distribution, the trapping state will be destroyed and \(P(n)\) will no longer be 0 for n$>$1, which makes \(g^{(2)} (0) > 0\) as we see from Fig. \ref{g2d3}. We can illustrate this behavior by a numerical example, for the case of a relative velocity spread of .01\%. We find in this case from the QTM simulation that \(P(n) > 0\) for \(n \le 7\) and \(\left\langle n \right\rangle  = 0.3918\) and \(\left\langle {n^2 } \right\rangle  = 0.4264\). These values yield an initial value of the correlation function \(g^{(2)}(0)\) to be 0.2254. It turns out that the wider the velocity distribution, the higher the value of \(g^{(2)}(0)\) till \(\Delta {\rm{v/v}}\) reaches \({\rm{0}}{\rm{.2\% }}\). When we increase the relative velocity spread beyond 0.2\%, the correlation is lost gradually due to the huge randomness in the interaction times between the atoms and the cavity, until a residual correlation persists starting from a relative velocity spread of 20\%. Fig. \ref{g2d4} demonstrates this behavior where we notice that correlation and hence the bunching of the cavity photons decreases gradually for relative velocity spreads of 0.2\%, 0.6\%, 1\%, 2\% and 20\%. 
As for the effect of velocity distribution on the correlation function of the total number of photons in the cavity (i.e., in the two modes combined), we observe,  as shown in Fig. \ref{g2d5}, a  behavior similar to the single mode correlation function described above for a velocity spread of 0.4\%, and \(g\tau _{{\mathop{\rm int}} }  = \frac{\pi }{{\sqrt 3 }}\) and \(\frac{{2\pi }}{{\sqrt 3 }}\).

\section{Conclusion}
We have applied the quantum trajectory method to the two-mode microlaser operating on $\Lambda$-type three level atoms.
We  verified  that  the  two-mode microlaser  does  reach  steady  state when  the  coupling 
between  the  atom  and  the  two  modes  is  not  symmetric,  the  case  where  the  detailed 
balance  condition is not guaranteed.
As  for  the  symmetric  operation  of  the microlaser, we 
explained the flat probability regions in the photon number probability distribution of any 
of  the  two  modes  and  emphasized  the  fact  that  the  existence  of  two  modes  equally 
coupled  to the atoms gives rise to these flat regions.
The super-poissonian distribution of 
the photon statistics of any of the two modes is a direct consequence of these flat regions 
and  causes  the  photon  correlation  to  be bunched  for most of  the  range of  the operating 
parameters  of  the  microlaser,  a  distinct  property  that  distinguishes  the  two-mode 
microlaser from the single mode microlaser. We introduced the analogy between the cavity field of  the  one-photon  trapping  state  and  the  resonance  fluorescence  radiation 
scattered from a  two-level atom. The correlation function of  this state  was fitted to a very 
simple  formula  for which we give  an  analytical  derivation in the Appendix.  Experimental 
measurement of  this anti-bunching  is considered a direct verification of  the quantized nature of  the 
field.  
 We investigated the effect of the atomic velocity spread on the statistical properties of the cavity field by 
the quantum trajectory method. We  found  that  the  trapping  states are very  sensitive  to 
the velocity  spread of  the atoms. The  phase transitions  in  the microlaser behavior are also 
destroyed by  the velocity distribution of  the atoms starting from  the  transitions at higher 
values of  the pumping parameter. The  second-order correlation is  in general   immune  to    
the  velocity  spread  of  the  atoms. For  the  special  case when  the microlaser  is  operating 
near  a  vacuum  trapping  state,  the  intensity  correlation  of  the  cavity  field  is  easily 
destroyed  by  a  relative  velocity  spread  as  low  as  0.2 \%.  Increasing  the  velocity  spread 
further does not affect  the  low  residual correlation  that survives  the velocity spread of  the 
atoms.    When  the  cavity  field  of  the  mono-velocity  atomic  beam  microlaser  exhibits  anti-bunching,  adding  randomness  to  the  atomic  velocities  counter-intuitively  increases  the amount of correlation. This noise-induced  correlation  stops at velocity  spreads 
of about 0.2 \%. Increasing the velocity spread further destroys the correlation gradually 
until a residual correlation persists for relative velocity spreads of 20\% and higher. These results should be useful to experimentalists interested in measuring the cavity field correlation function for real systems.

As a further investigation, we propose  developing  a  quantum  trajectory method  from  the Fokker-Planck equation of  the microlaser \cite{Filipowicz} and evolving coherent states of  radiation  instead  of  number  states,  as  the  coherent  states  are  the  closest  states  to  the classical radiation field.

\appendix
\section{}
In what follows we are going to derive equation (\ref{ftint}) in the text. We include the effect of spontaneous emission in the derivation and  show that under the conditions of strong coupling, compared to the spontaneous decay rate, and small interaction time compared to the lifetime of the excited state, the effect of spontaneous emission can be neglected. For simplicity,  in this derivation we ignore the effect of the cavity damping during the interaction time because it is negligible when the interaction time is very small compared to the cavity damping time \cite{meystre}.
To obtain the form of \(F\left( \tau  \right)\), we start by writing the interaction Hamiltonian of the atom-field system assuming that the transition between the two ground levels is forbidden:
\begin{equation}
H_{{\mathop{\rm int}} }  = \hbar g_1 \left( {a_1 ^ \dag  \sigma _1  + a_1 \sigma _1 ^ \dag  } \right) + \hbar g_2 \left( {a_2 ^ \dag  \sigma _2  + a_2 \sigma _2 ^ \dag  } \right),
\end{equation}
where $g_1$ and $g_2$ are the coupling strengths between the cavity field modes and the atomic dipoles of the two transitions, and $\sigma _1, \sigma _2$ are the atomic raising and lowering operators. 
It has to be emphasized here that \(g_1\) and \(g_2\) are in general complex quantities and 
the Hamiltonian should be written as: 
\(
H_{{\mathop{\rm int}} }  = \hbar  \left( g_1{a_1 ^ \dag  \sigma _1  + g_1^{*}a_1 \sigma _1 ^ \dag  } \right) + \hbar  \left( g_2{a_2 ^ \dag  \sigma _2  + g_2^{*}a_2 \sigma _2 ^ \dag  } \right)
\). However, since all quantities of interest will depend on \(\left |g_1  \right |^{2}\) and \(\left |g_2  \right |^{2}\), we will treat \(g_1\) and \(g_2\) as real positive quantities.
The interaction Hamiltonian between the atom and the bath of free space modes is given by a similar expression:
\begin{widetext}
\begin{equation}
\wp=\hbar \sum_k g_{1k} [a_k ^ \dag  \sigma _1 e^{i(\omega_1-\nu_k)t} + H.c] +\hbar \sum_k g_{2k} [a_k ^ \dag  \sigma _2 e^{i(\omega_2-\nu_k)t} + H.c]  	
\end{equation}
The Schrodinger's equation in the interaction picture, governing the evolution of the atom-field wavefunction is
\begin{equation}
i\hbar \frac{{\partial \left| \psi  \right\rangle }}{{\partial t}} = (H_{{\mathop{\rm int}} }+\wp) \left| \psi  \right\rangle, 
\end{equation}
where \(\left| {\psi (t)} \right\rangle\) is generally expressed as \cite{cui}:

	\[\left| {\psi (t)} \right\rangle  = \sum\limits_{n,m} c_{a,n,m} \left| {a,n,m,0_k} \right\rangle  + c_{b_1 ,n+1,m} \left| {b_1 ,n+1,m,0_k} \right\rangle  + c_{b_2 ,n,m+1} \left| {b_2 ,n,m+1,0_k}\right\rangle +
\]
\begin{equation}
\sum_k c_{b_1 ,n,m,1_k} \left| {b_1 ,n,m,1_k}\right\rangle +\sum_k c_{b_2 ,n,m,1_k} \left| {b_2 ,n,m,1_k} \right\rangle , 
\end{equation}
\end{widetext}

 and  \(n,m\) are the numbers of photons in mode 1 and mode 2 respectively. The last two terms represent the states corresponding to the occurence of a spontaneous emission into each of the two ground states. Higher order processes that involve exchange of photons between the cavity modes and the free space modes mediated by the atom are neglected. At the beginning of the interaction, the atom is in its excited state
\begin{equation}
\left| {\psi (0)} \right\rangle  = \sum\limits_{n,m} {c_{a,n,m} \left| {a,n,m} \right\rangle } 
\end{equation}
 and hence
\begin{equation}
P_{n,m} (0) = \left| {c_{a,n,m} } \right|^2. 
\end{equation}
Inserting \(\left|{\psi(t)} \right\rangle \) into Schrodinger's equation and applying the Weisskopf-Wigner theory of spontaneous emission \cite{scully} we obtain:
\begin{widetext}
\begin{equation}\label{d0}
	\dot c_{a,n,m}  = -i g_1 \sqrt {n + 1} c_{b_1 ,n + 1,m}  -i g_2 \sqrt {m + 1} c_{b_2 ,n,m + 1}-\frac{1}{2}\Gamma_a c_{a,n,m}, 
\end{equation}
\end{widetext}
\begin{equation}\label{d1}
	i\dot c_{b_1 ,n,m}  = g_1 \sqrt n c_{a,n - 1,m},
\end{equation}
 \begin{equation}\label{d2}
	i\dot c_{b_2 ,n,m}  = g_2 \sqrt m c_{a,n,m - 1}, 
\end{equation}

where the Weisskopf-Wigner integration over the free space modes has been performed and $\Gamma_a$ is the spontaneous decay rate. By combining Eq. (\ref{d0}, \ref{d1}, \ref{d2}) we obtain: 
 
\begin{equation}\label{ddt}
	\ddot c_{a,n,m}  =  - \left[ {g_1 ^2 \left( {n + 1} \right) + g_2 ^2 \left( {m + 1} \right)} \right]c_{a,n,m}-\frac{\Gamma_a }{2}\dot c_{a,n,m}. 
\end{equation}

By solving (\ref{d1}, \ref{d2}, \ref{ddt}) subjected to the initial conditions $c_{b_1,n,m}(0)=c_{b_2,n,m}(0)=0$  we get
\begin{widetext}
\begin{equation} \label{ca}
c_{a,n,m} (\tau _{{\mathop{\rm int}} } ) = e^{-\frac{\Gamma_a }{4}\tau _{int}} c_{a,n,m}(0) [\cos (g_{n,m} \tau_{int})-\frac{\Gamma_a }{4 g_{n,m}} \sin (g_{n,m} \tau_{int})],
\end{equation}
\end{widetext}
\begin{equation} \label{cb1}
	c_{b_1,n,m} (\tau _{{\mathop{\rm int}} } )=g_1\sqrt{n}\ e^{-\frac{\Gamma_a }{4}\tau _{int}}\frac{\sin (g_{n-1,m} \tau_{int})}{g_{n-1,m}}c_{a,n-1,m}(0),
\end{equation}
\begin{equation} \label{cb2}
	c_{b_2,n,m} (\tau _{{\mathop{\rm int}} } )=g_2\sqrt{m}\ e^{-\frac{\Gamma_a }{4}\tau _{int}}\frac{\sin (g_{n,m-1} \tau_{int})}{g_{n,m-1}}c_{a,n,m-1}(0),
\end{equation}
where 
\begin{equation}\label{g}
	g^2_{n,m}=\lambda^2(n,m)-\left(\frac{\Gamma_a }{4}\right)^2
\end{equation}

and $\lambda(n,m)$ is defined as
\begin{equation} \label{lambda}
	\lambda \left( {n,m} \right) = \sqrt {g_1 ^2 \left( {n + 1} \right) + g_2 ^2 \left( {m + 1} \right)}.
\end{equation}
\begin{widetext}
Probability conservation requires that
	\[	P_{n,m}(0)=\left |c_{a,n,m}(0)  \right |^2=\left | c_{a,n,m} (\tau _{{\mathop{\rm int}} } ) \right |^2+\left | c_{b_1,n+1,m} (\tau _{{\mathop{\rm int}} } ) \right |^2+\left | c_{b_2,n,m+1} (\tau _{{\mathop{\rm int}} } ) \right |^2+\sum_k \left | c_{b_1,n,m,1_k} (\tau _{{\mathop{\rm int}} } ) \right |^2+ 
\]
\begin{equation}
 \sum_k \left | c_{b_2,n,m,1_k} (\tau _{{\mathop{\rm int}} } ) \right |^2  
\end{equation}
 from which we can obtain the last two terms in the form

	\[\sum_k \left | c_{b_1,n,m,1_k} (\tau _{{\mathop{\rm int}} } ) \right |^2+ \sum_k \left | c_{b_2,n,m,1_k} (\tau _{{\mathop{\rm int}} } ) \right |^2=P_{n,m}(0)(1-e^{-\frac{\Gamma_a }{2}\tau _{int}}[\cos^2(g_{n,m}\tau _{int})+\frac{\Gamma_a }{4g_{n,m}}^2\sin^2(g_{n,m}\tau _{int})
\]
\begin{equation}\label{sum}
-\frac{\Gamma_a }{4g_{n,m}}\sin(2g_{n,m}\tau _{int})]-[g_1^2(n+1)+g_2^2(m+1)]e^{-\frac{\Gamma_a }{2}\tau _{int}}\frac{\sin^2(g_{n,m}\tau _{int})}{g_{n,m}}).
\end{equation}
At the same time, we have
 
\begin{equation}
	P_{n,m}(\tau _{int})=\left | c_{a,n,m} (\tau _{{\mathop{\rm int}} } ) \right |^2+\left | c_{b_1,n,m} (\tau _{{\mathop{\rm int}} } ) \right |^2+\left | c_{b_2,n,m} (\tau _{{\mathop{\rm int}} } ) \right |^2+\sum_k \left | c_{b_1,n,m,1_k} (\tau _{{\mathop{\rm int}} } ) \right |^2+  \sum_k \left | c_{b_2,n,m,1_k} (\tau _{{\mathop{\rm int}} } ) \right |^2 
\end{equation}
which can be rewritten in terms of (\ref{ca}, \ref{cb1}, \ref{cb2}) as 
	\[P_{n,m}(\tau _{int})=P_{n,m}(0)e^{-\frac{\Gamma_a }{2}\tau _{int}}[\cos^2(g_{n,m}\tau _{int})+\frac{\Gamma_a }{4g_{n,m}}\sin^2(g_{n,m}\tau _{int})-\frac{\Gamma_a }{4g_{n,m}}\sin(2g_{n,m}\tau _{int})]
\]
	\[	+ g_1^2\ n\ e^{-\frac{\Gamma_a }{2}\tau _{int}}\frac{\sin^2(g_{n-1,m}\tau _{int})}{g_{n-1,m}}P_{n-1,m}(0)
+g_2^2\ m\ e^{-\frac{\Gamma_a }{2}\tau _{int}}\frac{\sin^2(g_{n,m-1}\tau _{int})}{g_{n,m-1}}P_{n,m-1}(0) 
\]
\begin{equation} \label{pt}
+\sum_k \left | c_{b_1,n,m,1_k} (\tau _{{\mathop{\rm int}} } ) \right |^2+ \sum_k \left | c_{b_2,n,m,1_k} (\tau _{{\mathop{\rm int}} } ) \right |^2. 
\end{equation}

It can be shown by straight algebra that applying the two conditions: $\Gamma_a \tau_{int} <<1$ and \(g_{1,2} > \Gamma_a \) into Eq.( \ref{g}, \ref{sum}, \ref{pt}) will lead to

	\[	P_{n,m} (\tau _{{\mathop{\rm int}} } ) = P_{n,m} (0)cos^2 \left[ {\lambda \left( {n,m} \right)\tau _{{\mathop{\rm int}} } } \right] + g_1^2 n P_{n - 1,m} (0)\frac{{\sin ^2 \left[ {\lambda \left( {n - 1,m} \right)\tau _{{\mathop{\rm int}} } } \right]}}{{\lambda ^2 \left( {n - 1,m} \right)}}
\]
\begin{equation}	 
+ g_2^2 m P_{n,m - 1} (0)\frac{{\sin ^2 \left[ {\lambda \left( {n,m - 1} \right)\tau _{{\mathop{\rm int}} } } \right]}}{{\lambda ^2 \left( {n,m - 1} \right)}}
\end{equation}
which is the same equation used in the text.
\end{widetext}

\section{}
We are going to prove that for the one photon trapping state, \(g^{(2)} (\tau )\) is exactly equal to 
\(f(\tau )\) given in (\ref{ft}) for the case of the two level single mode microlaser. Starting from the definition of \(g^{(2)} (\tau )\) for a quantized field:
\begin{equation}
g^{(2)} (\tau ) = \frac{{\left\langle {a^ \dag  (0)a^ \dag  (\tau )a(\tau )a(0)} \right\rangle }}{{\left\langle n \right\rangle ^2}}.
\end{equation}
where \(\tau  = 0\) represents the steady state. For the one photon trapping state only the states 
\(\left| 0 \right\rangle ,\left| 1 \right\rangle\) are accessible to the cavity field. We can then write

\begin{eqnarray*}
	g^{(2)} (\tau ) = \frac{{\sum\limits_{n = 0,1} {\left\langle n \right|a^\dag  (0)a^ \dag  (\tau )a(\tau )a(0)\left| n \right\rangle } }}{{\left\langle n \right\rangle ^2 }}
\\= \frac{{\sum\limits_{n = 0,1} {n.\left\langle {n - 1} \right|a^ \dag  (\tau )a(\tau )\left| {n - 1} \right\rangle } }}{{\left\langle n \right\rangle ^2 }}
\end{eqnarray*}
\begin{eqnarray*}
	= \frac{{\left\langle 0 \right|a^\dag  (\tau )a(\tau )\left| 0 \right\rangle }}{{\left\langle n \right\rangle ^2 }}
	\end{eqnarray*}
	\begin{equation}\label{gg20}
	 = \frac{{\left\langle 0 \right|\hat n(\tau )\left| 0 \right\rangle }}{{\left\langle n \right\rangle ^2 }},\\
\end{equation}

where \(\hat n(\tau )\) is the number operator \(a^\dag  (\tau )a(\tau )\).  Let's calculate the evolution of the mean number of photons $n(t)$ in the Schrodinger's picture, letting $a^ \dag  (0)=a^ \dag$ and $a(0)=a$. Under the one-photon trapping state condition, we have only two possibilities: to find 0 or 1 photon inside the cavity and hence 
\begin{equation}\label{mn}
	n(t) = \sum\limits_n {n.p_n } (t) = p_1 (t),
\end{equation}
 since \(p_n (t) = 0\) for \(n \ne 0,1\).  It can be shown from the single mode microlaser master equation (see for example \cite{Quang}) that evolution of \(p_1 (t)\) is governed by:
\begin{equation}
\label{p1}
	\dot p_1 (t) = p_0 (t)R\sin ^2 \left[ {g\tau _{{\mathop{\rm int}} } \sqrt 1 } \right] - \gamma p_1 (t).
\end{equation}
We should note that the $\sqrt 1$ factor in (\ref{p1}) that will be dropped in the following becomes $\sqrt 2$ in the two-mode case.
From (\ref{p1}, \ref{mn}), we deduce that the evolution of the mean number of photons inside the cavity in the one photon trapping state is governed by 

\begin{equation}\label{ndot}
	\dot n(t) = p_0 (t)R\sin ^2 \left[ {g\tau _{{\mathop{\rm int}} }  } \right] - \gamma n(t).
\end{equation}

This equation is intuitive for the one-photon trapping state and could have been written directly without referring to the master equation. The second term on the right hand side represents the rate of photon loss from the cavity while the first term represents the number of photons injected inside the cavity per unit time. 
At steady state, \(\dot p_1 (0) = 0\) and hence  
\[
0 = \left\{ {1 - p_1 (0)} \right\}R\sin ^2 \left[ {g\tau _{{\mathop{\rm int}} }  } \right] - \gamma p_1 (0),
\]
which leads to 
\begin{equation}\label{nn2}
	\left\langle n \right\rangle  = \left\langle {n^2 } \right\rangle  = p_1 (0) = \frac{{R\sin ^2 \left[ {g\tau _{{\mathop{\rm int}} }  } \right]}}{{\gamma  + R\sin ^2 \left[ {g\tau _{{\mathop{\rm int}} }  } \right]}}.
\end{equation}

Since \(p_1 (t) + p_0 (t) = 1\), we can write (\ref{ndot}) as: 
\[
\dot n(t) = \left( {1 - n(t)} \right)R\sin ^2 \left[ {g\tau _{{\mathop{\rm int}} }  } \right] - \gamma n(t)
\]
\[
 = R\sin ^2 \left[ {g\tau _{{\mathop{\rm int}} } } \right] - \left( {R\sin ^2 \left[ {g\tau _{{\mathop{\rm int}} }  } \right] + \gamma } \right)n(t)
\]

\begin{equation}\label{ab}
	 = A - Bn(t),
\end{equation}
where 
\begin{equation}\label{a}
	A = R\sin ^2 \left[ {g\tau _{{\mathop{\rm int}} }  } \right],B = \left( {R\sin ^2 \left[ {g\tau _{{\mathop{\rm int}} } } \right] + \gamma } \right).
\end{equation}
By integrating (\ref{ab}) from 
\(0 \to \tau \) we get 
\begin{equation}
	\int\limits_{n(0)}^{n(\tau )} {\frac{{dn}}{{A - B.n}}}  = \int\limits_0^\tau  {dt}.
\end{equation}

By solving for \(n(\tau )\), we obtain: 
\begin{equation}\label{n1}
	n(\tau ) = \frac{A}{B} - \left[ {\frac{A}{B} - n(0)} \right]e^{ - B\tau}.
\end{equation}
 
 Back to the Heisenberg picture where the operators vary in time, we have:

\begin{equation}\label{n2}
	n(0) = \sum\limits_n {\left\langle n \right|\hat n(0)} \left| n \right\rangle, 
\end{equation}

\begin{equation}\label{n3}
n(\tau ) = \sum\limits_n {\left\langle n \right|\hat n(\tau )} \left| n \right\rangle, 
\end{equation}

\begin{equation}\label{n4}
	\hat n(0) = \sum\limits_n {\left\langle n \right|\hat n(0)} \left| n \right\rangle \left| n \right\rangle \left\langle n \right|=a^ \dag a,
\end{equation}

and by assuming that \(\hat n(\tau )\) is diagonal in the form:

\begin{equation}\label{n5}
	\hat n(\tau ) = \sum\limits_n {\left\langle n \right|\hat n(\tau )} \left| n \right\rangle \left| n \right\rangle \left\langle n \right|,
\end{equation}

one may infer from (\ref{n1}), (\ref{n2}), (\ref{n3}), (\ref{n4}) and (\ref{n5}) that 

\begin{equation}
	\hat n(\tau ) = \frac{A}{B} - \left[ {\frac{A}{B} - \hat n(0)} \right]e^{ - B\tau }, 
\end{equation}

and hence the numerator of (\ref{gg20}) equals 
\begin{equation}\label{0n0}
	\left\langle 0 \right|\hat n(\tau )\left| 0 \right\rangle  = \frac{A}{B} - \left[ {\frac{A}{B} - 0} \right]e^{ - B\tau } = \frac{A}{B}\left( {1 - e^{ - B\tau } } \right).
\end{equation}

From (\ref{gg20}), (\ref{nn2}), (\ref{a}) and (\ref{0n0}) we can write 
\begin{equation}
g^{(2)} (\tau ) = \left( {1 - e^{ - B\tau } } \right),
\end{equation}
which is the same as equation (\ref{ft}). When this proof is generalized to the
two-mode case, B will be expressed as $B = {R\sin ^2 \left[ {g\tau _{{\mathop{\rm int}} } }\sqrt 2 \right] + \gamma }$.

\begin{acknowledgments}
This work was supported by King Fahd University of Petroleum and Minerals (KFUPM). The comments of  Jorg Evers are highly appreciated.
\end{acknowledgments}
%\bibliography{references}

\begin{thebibliography}{29}
\expandafter\ifx\csname natexlab\endcsname\relax\def\natexlab#1{#1}\fi
\expandafter\ifx\csname bibnamefont\endcsname\relax
  \def\bibnamefont#1{#1}\fi
\expandafter\ifx\csname bibfnamefont\endcsname\relax
  \def\bibfnamefont#1{#1}\fi
\expandafter\ifx\csname citenamefont\endcsname\relax
  \def\citenamefont#1{#1}\fi
\expandafter\ifx\csname url\endcsname\relax
  \def\url#1{\texttt{#1}}\fi
\expandafter\ifx\csname urlprefix\endcsname\relax\def\urlprefix{URL }\fi
\providecommand{\bibinfo}[2]{#2}
\providecommand{\eprint}[2][]{\url{#2}}

\bibitem[{\citenamefont{An et~al.}(1994)\citenamefont{An, Childs, Dasari, and
  Feld}}]{An}
\bibinfo{author}{\bibfnamefont{K.}~\bibnamefont{An}},
  \bibinfo{author}{\bibfnamefont{J.~J.} \bibnamefont{Childs}},
  \bibinfo{author}{\bibfnamefont{R.~R.} \bibnamefont{Dasari}},
  \bibnamefont{and} \bibinfo{author}{\bibfnamefont{M.~S.} \bibnamefont{Feld}},
  \bibinfo{journal}{Phys. Rev. Lett.} \textbf{\bibinfo{volume}{73}},
  \bibinfo{pages}{3375} (\bibinfo{year}{1994}).

\bibitem[{\citenamefont{{Maddox}}(1995)}]{Mad}
\bibinfo{author}{\bibfnamefont{J.}~\bibnamefont{{Maddox}}},
  \bibinfo{journal}{\nat} \textbf{\bibinfo{volume}{373}}, \bibinfo{pages}{101}
  (\bibinfo{year}{1995}).

\bibitem[{\citenamefont{Choi et~al.}(2006)\citenamefont{Choi, Lee, An,
  Fang-Yen, Dasari, and Feld}}]{Choi}
\bibinfo{author}{\bibfnamefont{W.}~\bibnamefont{Choi}},
  \bibinfo{author}{\bibfnamefont{J.-H.} \bibnamefont{Lee}},
  \bibinfo{author}{\bibfnamefont{K.}~\bibnamefont{An}},
  \bibinfo{author}{\bibfnamefont{C.}~\bibnamefont{Fang-Yen}},
  \bibinfo{author}{\bibfnamefont{R.~R.} \bibnamefont{Dasari}},
  \bibnamefont{and} \bibinfo{author}{\bibfnamefont{M.~S.} \bibnamefont{Feld}},
  \bibinfo{journal}{Phys. Rev. Lett.} \textbf{\bibinfo{volume}{96}},
  \bibinfo{pages}{093603} (\bibinfo{year}{2006}).

\bibitem[{\citenamefont{Fang-Yen et~al.}(2006)\citenamefont{Fang-Yen, Yu, Ha,
  Choi, An, Dasari, and Feld}}]{pra2006}
\bibinfo{author}{\bibfnamefont{C.}~\bibnamefont{Fang-Yen}},
  \bibinfo{author}{\bibfnamefont{C.~C.} \bibnamefont{Yu}},
  \bibinfo{author}{\bibfnamefont{S.}~\bibnamefont{Ha}},
  \bibinfo{author}{\bibfnamefont{W.}~\bibnamefont{Choi}},
  \bibinfo{author}{\bibfnamefont{K.}~\bibnamefont{An}},
  \bibinfo{author}{\bibfnamefont{R.~R.} \bibnamefont{Dasari}},
  \bibnamefont{and} \bibinfo{author}{\bibfnamefont{M.~S.} \bibnamefont{Feld}},
  \bibinfo{journal}{Phys. Rev. A} \textbf{\bibinfo{volume}{73}},
  \bibinfo{pages}{041802} (\bibinfo{year}{2006}).

\bibitem[{\citenamefont{Hong et~al.}(2009)\citenamefont{Hong, Seo, Lee, Song,
  Choi, Fang-Yen, Dasari, Feld, Lee, and An}}]{pra2009}
\bibinfo{author}{\bibfnamefont{H.-G.} \bibnamefont{Hong}},
  \bibinfo{author}{\bibfnamefont{W.}~\bibnamefont{Seo}},
  \bibinfo{author}{\bibfnamefont{M.}~\bibnamefont{Lee}},
  \bibinfo{author}{\bibfnamefont{Y.}~\bibnamefont{Song}},
  \bibinfo{author}{\bibfnamefont{W.}~\bibnamefont{Choi}},
  \bibinfo{author}{\bibfnamefont{C.}~\bibnamefont{Fang-Yen}},
  \bibinfo{author}{\bibfnamefont{R.~R.} \bibnamefont{Dasari}},
  \bibinfo{author}{\bibfnamefont{M.~S.} \bibnamefont{Feld}},
  \bibinfo{author}{\bibfnamefont{J.-H.} \bibnamefont{Lee}}, \bibnamefont{and}
  \bibinfo{author}{\bibfnamefont{K.}~\bibnamefont{An}}, \bibinfo{journal}{Phys.
  Rev. A} \textbf{\bibinfo{volume}{79}}, \bibinfo{pages}{033816}
  (\bibinfo{year}{2009}).

\bibitem[{\citenamefont{Yu et~al.}(2001)\citenamefont{Yu, Fang-Yen, Aljalal,
  Dasari, An, , and Feld}}]{book}
\bibinfo{author}{\bibfnamefont{C.}~\bibnamefont{Yu}},
  \bibinfo{author}{\bibfnamefont{C.}~\bibnamefont{Fang-Yen}},
  \bibinfo{author}{\bibfnamefont{A.}~\bibnamefont{Aljalal}},
  \bibinfo{author}{\bibfnamefont{R.}~\bibnamefont{Dasari}},
  \bibinfo{author}{\bibfnamefont{K.}~\bibnamefont{An}}, , \bibnamefont{and}
  \bibinfo{author}{\bibfnamefont{M.~S.} \bibnamefont{Feld}}, in
  \emph{\bibinfo{booktitle}{McGraw-Hill Yearbook of Science and Technology}}
  (\bibinfo{publisher}{McGraw-Hill}, \bibinfo{year}{2001}).

\bibitem[{Mes()}]{Mess}
\bibinfo{note}{For a review of the Interaction of bimodal fields with few-level
  atoms in cavities, see: A. Messina a, S. Maniscalco, and A. Napoli, J. Mod.
  Opt., 50, 1 (2003).}

\bibitem[{\citenamefont{Kien et~al.}(1995)\citenamefont{Kien, Meyer, Rathe,
  Scully, Walther, and Zhu}}]{Fam95}
\bibinfo{author}{\bibfnamefont{F.~L.} \bibnamefont{Kien}},
  \bibinfo{author}{\bibfnamefont{G.~M.} \bibnamefont{Meyer}},
  \bibinfo{author}{\bibfnamefont{U.~W.} \bibnamefont{Rathe}},
  \bibinfo{author}{\bibfnamefont{M.~O.} \bibnamefont{Scully}},
  \bibinfo{author}{\bibfnamefont{H.}~\bibnamefont{Walther}}, \bibnamefont{and}
  \bibinfo{author}{\bibfnamefont{S.-Y.} \bibnamefont{Zhu}},
  \bibinfo{journal}{Phys. Rev. A} \textbf{\bibinfo{volume}{52}},
  \bibinfo{pages}{3279} (\bibinfo{year}{1995}).

\bibitem[{\citenamefont{Dalibard et~al.}(1992)\citenamefont{Dalibard, Castin,
  and M\o{}lmer}}]{Dal}
\bibinfo{author}{\bibfnamefont{J.}~\bibnamefont{Dalibard}},
  \bibinfo{author}{\bibfnamefont{Y.}~\bibnamefont{Castin}}, \bibnamefont{and}
  \bibinfo{author}{\bibfnamefont{K.}~\bibnamefont{M\o{}lmer}},
  \bibinfo{journal}{Phys. Rev. Lett.} \textbf{\bibinfo{volume}{68}},
  \bibinfo{pages}{580} (\bibinfo{year}{1992}).

\bibitem[{\citenamefont{Dum et~al.}(1992)\citenamefont{Dum, Parkins, Zoller,
  and Gardiner}}]{Dum}
\bibinfo{author}{\bibfnamefont{R.}~\bibnamefont{Dum}},
  \bibinfo{author}{\bibfnamefont{A.~S.} \bibnamefont{Parkins}},
  \bibinfo{author}{\bibfnamefont{P.}~\bibnamefont{Zoller}}, \bibnamefont{and}
  \bibinfo{author}{\bibfnamefont{C.~W.} \bibnamefont{Gardiner}},
  \bibinfo{journal}{Phys. Rev. A} \textbf{\bibinfo{volume}{46}},
  \bibinfo{pages}{4382} (\bibinfo{year}{1992}).

\bibitem[{\citenamefont{Cresser and Pickles}(1996)}]{Cres}
\bibinfo{author}{\bibfnamefont{J.~D.} \bibnamefont{Cresser}} \bibnamefont{and}
  \bibinfo{author}{\bibfnamefont{S.~M.} \bibnamefont{Pickles}},
  \bibinfo{journal}{Quantum and Semiclassical Optics: Journal of the European
  Optical Society Part B} \textbf{\bibinfo{volume}{8}}, \bibinfo{pages}{73}
  (\bibinfo{year}{1996}).

\bibitem[{\citenamefont{Carmichael}(1993)}]{Car}
\bibinfo{author}{\bibfnamefont{H.~J.} \bibnamefont{Carmichael}},
  \emph{\bibinfo{title}{An Open Systems Approach to Quantum Optics}}, Lecture
  Notes in Physics (\bibinfo{publisher}{Springer Verlag},
  \bibinfo{address}{Berlin}, \bibinfo{year}{1993}).

\bibitem[{\citenamefont{Filipowicz et~al.}(1986)\citenamefont{Filipowicz,
  Javanainen, and Meystre}}]{Filipowicz}
\bibinfo{author}{\bibfnamefont{P.}~\bibnamefont{Filipowicz}},
  \bibinfo{author}{\bibfnamefont{J.}~\bibnamefont{Javanainen}},
  \bibnamefont{and} \bibinfo{author}{\bibfnamefont{P.}~\bibnamefont{Meystre}},
  \bibinfo{journal}{Phys. Rev. A} \textbf{\bibinfo{volume}{34}},
  \bibinfo{pages}{3077} (\bibinfo{year}{1986}).

\bibitem[{\citenamefont{Kien et~al.}(1994)\citenamefont{Kien, Meyer, Scully,
  Walther, and Zhu}}]{Fam94}
\bibinfo{author}{\bibfnamefont{F.~L.} \bibnamefont{Kien}},
  \bibinfo{author}{\bibfnamefont{G.~M.} \bibnamefont{Meyer}},
  \bibinfo{author}{\bibfnamefont{M.~O.} \bibnamefont{Scully}},
  \bibinfo{author}{\bibfnamefont{H.}~\bibnamefont{Walther}}, \bibnamefont{and}
  \bibinfo{author}{\bibfnamefont{S.-Y.} \bibnamefont{Zhu}},
  \bibinfo{journal}{Phys. Rev. A} \textbf{\bibinfo{volume}{49}},
  \bibinfo{pages}{1367} (\bibinfo{year}{1994}).

\bibitem[{\citenamefont{Scully and Zubairy}(1997)}]{scully}
\bibinfo{author}{\bibfnamefont{M.}~\bibnamefont{Scully}} \bibnamefont{and}
  \bibinfo{author}{\bibfnamefont{M.~S.} \bibnamefont{Zubairy}},
  \emph{\bibinfo{title}{Quantum Optics}} (\bibinfo{publisher}{Cambridge
  University Press}, \bibinfo{year}{1997}).

\bibitem[{\citenamefont{Meystre and III}(1991)}]{meystre}
\bibinfo{author}{\bibfnamefont{P.}~\bibnamefont{Meystre}} \bibnamefont{and}
  \bibinfo{author}{\bibfnamefont{M.~S.} \bibnamefont{III}},
  \emph{\bibinfo{title}{Elements of Quantum Optics}}
  (\bibinfo{publisher}{Springer Verlag}, \bibinfo{address}{Berlin},
  \bibinfo{year}{1991}).

\bibitem[{\citenamefont{An"}(1995)}]{thesis}
\bibinfo{author}{\bibfnamefont{K.}~\bibnamefont{An"}}, Ph.D. thesis,
  \bibinfo{school}{Massachusetts Institute of Technology}
  (\bibinfo{year}{1995}), \eprint{arXiv:physics/0412181v1}.

\bibitem[{\citenamefont{Yen"}(2002)}]{Fangyen}
\bibinfo{author}{\bibfnamefont{C.~F.} \bibnamefont{Yen"}}, Ph.D. thesis,
  \bibinfo{school}{Massachusetts Institute of Technology}
  (\bibinfo{year}{2002}), \eprint{arXiv:physics/0412181v1}.

\bibitem[{\citenamefont{Thomsen}(1953)}]{Thomsen}
\bibinfo{author}{\bibfnamefont{J.~S.} \bibnamefont{Thomsen}},
  \bibinfo{journal}{Phys. Rev.} \textbf{\bibinfo{volume}{91}},
  \bibinfo{pages}{1263} (\bibinfo{year}{1953}).

\bibitem[{\citenamefont{Roa}(1994)}]{comment1}
\bibinfo{author}{\bibfnamefont{L.}~\bibnamefont{Roa}}, \bibinfo{journal}{Phys.
  Rev. A} \textbf{\bibinfo{volume}{50}}, \bibinfo{pages}{R1995}
  (\bibinfo{year}{1994}).

\bibitem[{\citenamefont{Arun et~al.}(2000)\citenamefont{Arun, Agarwal, Scully,
  and Walther}}]{comment2}
\bibinfo{author}{\bibfnamefont{R.}~\bibnamefont{Arun}},
  \bibinfo{author}{\bibfnamefont{G.~S.} \bibnamefont{Agarwal}},
  \bibinfo{author}{\bibfnamefont{M.~O.} \bibnamefont{Scully}},
  \bibnamefont{and} \bibinfo{author}{\bibfnamefont{H.}~\bibnamefont{Walther}},
  \bibinfo{journal}{Phys. Rev. A} \textbf{\bibinfo{volume}{62}},
  \bibinfo{pages}{023809} (\bibinfo{year}{2000}).

\bibitem[{\citenamefont{\ifmmode~\check{S}\else \v{S}\fi{}karja
  et~al.}(1999)\citenamefont{\ifmmode~\check{S}\else \v{S}\fi{}karja,
  Manko\ifmmode \check{c}\else \v{c}\fi{} Bor\ifmmode~\check{s}\else
  \v{s}\fi{}tnik, L\"offler, and Walther}}]{comment3}
\bibinfo{author}{\bibfnamefont{M.}~\bibnamefont{\ifmmode~\check{S}\else
  \v{S}\fi{}karja}},
  \bibinfo{author}{\bibfnamefont{N.}~\bibnamefont{Manko\ifmmode \check{c}\else
  \v{c}\fi{} Bor\ifmmode~\check{s}\else \v{s}\fi{}tnik}},
  \bibinfo{author}{\bibfnamefont{M.}~\bibnamefont{L\"offler}},
  \bibnamefont{and} \bibinfo{author}{\bibfnamefont{H.}~\bibnamefont{Walther}},
  \bibinfo{journal}{Phys. Rev. A} \textbf{\bibinfo{volume}{60}},
  \bibinfo{pages}{3229} (\bibinfo{year}{1999}).

\bibitem[{\citenamefont{Arun and Agarwal}(2002)}]{comment4}
\bibinfo{author}{\bibfnamefont{R.}~\bibnamefont{Arun}} \bibnamefont{and}
  \bibinfo{author}{\bibfnamefont{G.~S.} \bibnamefont{Agarwal}},
  \bibinfo{journal}{Phys. Rev. A} \textbf{\bibinfo{volume}{66}},
  \bibinfo{pages}{043812} (\bibinfo{year}{2002}).

\bibitem[{\citenamefont{Loudon}(2000)}]{loudon}
\bibinfo{author}{\bibfnamefont{R.}~\bibnamefont{Loudon}},
  \emph{\bibinfo{title}{The Quantum Theory of Light}}
  (\bibinfo{publisher}{Oxford University Press}, \bibinfo{year}{2000}).

\bibitem[{\citenamefont{Quang}(1992)}]{Quang}
\bibinfo{author}{\bibfnamefont{T.}~\bibnamefont{Quang}},
  \bibinfo{journal}{Phys. Rev. A} \textbf{\bibinfo{volume}{46}},
  \bibinfo{pages}{682} (\bibinfo{year}{1992}).

\bibitem[{\citenamefont{Aljalal"}(2001)}]{aljalal}
\bibinfo{author}{\bibfnamefont{A.}~\bibnamefont{Aljalal"}}, Ph.D. thesis,
  \bibinfo{school}{Massachusetts Institute of Technology}
  (\bibinfo{year}{2001}).

\bibitem[{\citenamefont{Weidinger et~al.}(1999)\citenamefont{Weidinger, Varcoe,
  Heerlein, and Walther}}]{walther}
\bibinfo{author}{\bibfnamefont{M.}~\bibnamefont{Weidinger}},
  \bibinfo{author}{\bibfnamefont{B.~T.~H.} \bibnamefont{Varcoe}},
  \bibinfo{author}{\bibfnamefont{R.}~\bibnamefont{Heerlein}}, \bibnamefont{and}
  \bibinfo{author}{\bibfnamefont{H.}~\bibnamefont{Walther}},
  \bibinfo{journal}{Phys. Rev. Lett.} \textbf{\bibinfo{volume}{82}},
  \bibinfo{pages}{3795} (\bibinfo{year}{1999}).

\bibitem[{\citenamefont{Rekdal and Skagerstam}(2002)}]{Rekdal}
\bibinfo{author}{\bibfnamefont{P.~K.} \bibnamefont{Rekdal}} \bibnamefont{and}
  \bibinfo{author}{\bibfnamefont{B.-S.~K.} \bibnamefont{Skagerstam}},
  \bibinfo{journal}{Physica A: Statistical Mechanics and its Applications}
  \textbf{\bibinfo{volume}{305}}, \bibinfo{pages}{404 } (\bibinfo{year}{2002}).

\bibitem[{\citenamefont{Cui and Raymer}(2005)}]{cui}
\bibinfo{author}{\bibfnamefont{G.}~\bibnamefont{Cui}} \bibnamefont{and}
  \bibinfo{author}{\bibfnamefont{M.}~\bibnamefont{Raymer}},
  \bibinfo{journal}{Opt. Express} \textbf{\bibinfo{volume}{13}},
  \bibinfo{pages}{9660} (\bibinfo{year}{2005}).

\end{thebibliography}

\end{document}